\documentclass[10pt]{article}
\usepackage{amsmath}
\usepackage{amsthm}
\usepackage{amsfonts}
\usepackage{amssymb}
\usepackage{url}
\usepackage{eucal}
\usepackage{mathrsfs}
\usepackage{graphicx,graphics}
\usepackage[T1]{fontenc}%
\usepackage{url}
\usepackage{epsfig}
\usepackage{epstopdf}
\usepackage{xcolor}
\usepackage[english]{babel}

\newcommand\fig[1] {{\rm Figure}~\ref{fig:#1}}
\newcommand\figs[1] {~\ref{fig:#1}}
\newcommand\labfig[1] {\label{fig:#1}}

\newcommand{\bfm}[1]{\mbox{\boldmath ${#1}$}}
\newcommand{\nonum}{\nonumber \\}
\newcommand\eq[1] {(\ref{#1})}
\newcommand{\beqa}{\begin{eqnarray}}
\newcommand{\eeqa}[1]{\label{#1}\end{eqnarray}}
\newcommand{\beq}{\begin{equation}}
\newcommand{\eeq}[1]{\label{#1}\end{equation}}

\newcommand{\Grad}{\nabla}
\newcommand{\Div}{\nabla \cdot}

\newcommand{\Md}{\partial}

\newcommand{\Ga}{\alpha}
\newcommand{\Gb}{\beta}
\newcommand{\Gd}{\delta}
\newcommand{\Ge}{\epsilon}

\newcommand{\Gg}{\gamma}
\newcommand{\Gc}{\chi}

\newcommand{\BGve}{\bfm\varepsilon}

\newcommand{\BGs}{\bfm\sigma}

\newcommand{\CP}{{\cal P}}

\newcommand{\CT}{{\cal T}}

\newcommand{\bpm}{\begin{pmatrix}}
\newcommand{\epm}{\end{pmatrix}}

\def\b0{\bf 0}

\def\Bd{{\bf d}}
\def\Be{{\bf e}}

\def\Bj{{\bf j}}

\def\Bm{{\bf m}}
\def\Bn{{\bf n}}

\def\Bu{{\bf u}}

\def\Bx{{\bf x}}
\def\By{{\bf y}}

\def\BJ{{\bf J}}

\def\BV{{\bf V}}

\usepackage{caption,subcaption}
\begin{document}
\vspace{-1in}
\title{Field Patterns: A New Mathematical Object}
\author{Graeme W. Milton and Ornella Mattei\\
	\small{Department of Mathematics, University of Utah, Salt Lake City UT 84112, USA}}
\date{}
\maketitle
\vskip 1.cm
\begin{abstract}
Field patterns occur in space-time microstructures such that a disturbance propagating along a characte-ristic line, does not evolve into a cascade of disturbances, but rather concentrates on a pattern of characteristic lines. This pattern is the field pattern. In one spatial direction plus time, the field patterns occur when the slope of the characteristics is, in a sense, commensurate with the space-time microstructure. Field patterns with different spatial shifts do not generally interact, but rather evolve as if they live in separate dimensions, as many dimensions as the number of field patterns. Alternatively
one can view a collection as a multicomponent potential, with as many components as the number of field patterns. Presumably if one added a tiny nonlinear term to the wave equation one would then see interactions between these field patterns in the multidimensional space that one can consider them to live, or between the different field components of the multicomponent potential if one views them that way. As a result of $\CP$$\CT$--symmetry many of the complex eigenvalues of an appropriately defined transfer matrix have unit norm and hence the corresponding eigenvectors correspond to propagating modes. There are also modes that blow up exponentially with time.
\end{abstract}

\section{Introduction}
\setcounter{equation}{0}

Here we introduce the theory of field patterns. Field patterns develop when waves, concentrated on characteristic lines, interact with certain special space-time microstructures. They occur when the space-time microstructure is such that the propagation along characteristics does not develop into a complicated 
cascade of space-time lines, but rather concentrates along particular patterns: these are the field patterns. There is an obvious connection with dynamical systems.

What makes field patterns mathematically novel is their appearance in wave equations and the
associated multidimensional, or multicomponent, character of a collection of field patterns even though the wave equation is, say, a scalar wave-equation in 1-, 2-, or 3-dimensions plus time.
Here for simplicity we will focus on the appearance of field patterns in scalar wave equations in $1$ spatial dimension plus time. 

The field patterns we introduce here can have both a particle-like aspect, as the patterns
are concentrated on lines in space-time, and a wave-like aspect, as the patterns at long times can develop
wave-like features. This hints of a connection with quantum mechanics. Another connection is the multidimensional nature of field patterns. In the 
noninteracting model of field patterns developed here,
different field patterns in a collection, while occupying the same space-time continuum, act as if they live
in separate dimensions. The multidimensional nature of collections of field patterns is quite different to that associated with multiscale homogenization
theory. In multiscale homogenization, with say periodicity at all but the largest length scale,  the moduli of the materials and the fields are often modeled 
in $d$-dimensions by
functions of the form
\beq u_\Ge(\Bx)=f(\Bx,\Bx/\Ge,\Bx/\Ge^2,\Bx/\Ge^3,\ldots,\Bx/\Ge^{m-1}), \eeq{0.0}
where $\Ge>0$ is a very small parameter giving the ratio between length scales and $f(\Bx,\By_1,\By_2,\By_3,\ldots,\By_{m-1})$ is a function living in a $md$ dimensional space, that is periodic in each of the variables
$\By_1,\By_2,\ldots,\By_m$, but not necessarily periodic in $\Bx$ 
(see, for example, \cite{Bensoussan:1978:AAP,Attouch:1984:VCF,Zhikov:1994:HDO,Tartar:2009:GTH}).  
Conversely, the notions of two-scale and multiscale convergence \cite{Nguetseng:1989:GCR,Allaire:1992:HTS} allow one
to go from sequences of functions  $u_\Ge(\Bx)$ parametrized by $\Ge$ to the multidimensional function $f(\Bx,\By_1,\By_2,\By_3,\ldots,\By_{m-1})$.
In this multiscale homogenization, the different dimensions $\Bx$, $\By_1,\By_2$, $\By_3$,$\ldots,$$\By_{m-1}$ play very different roles, each associated
with a different length scale. By contrast, the different dimensions associated with a collection of field patterns play more or less an equivalent role,
as do the different dimensions associated with say the multielectron Schr{\"o}dinger equation. Rather than using
a multidimensional space, an alternative way of viewing noninteracting field patterns is as a multicomponent 
potential, where each component potential is associated with one field pattern.

We do not study models where there are interactions between the field patterns. This would
require adding non-linear terms to the wave equation. Without such terms, a superposition of two field pattern solutions is 
also a solution: the field patterns do not interact. 

Some quantum mechanical aspects are notably absent from the models we study here: significantly, there is nothing analogous to the collapse of the wave-function; there is no stochastic element in the evolutions we describe here; and moreover there is no quantization of the solutions.
It is curious that some of these quantum mechanical aspects have been seen in the dynamics of "walking droplets" 
of silicon oil on a vibrating silicon oil fluid surface, that may be interpreted as a space-time microstructure \cite{Couder:2005:DPW,Couder:2006:SPD,Bush:2015:PWH}. In a certain region of instability, these droplets start "walking", creating their own "pilot wave" on the surface. The trajectory of the droplet can seem stochastic as the motion of a droplet after a bounce depends on the orientation of the fluid surface where it lands, and this can appear quite random: nevertheless the ensemble averaged statistics exhibit wave-like features. The "holy-grail" would be that a complete explanation of quantum mechanics emerges from a combination of these "pilot wave" ideas, and the ideas in the theory of field patterns introduced here. The bold conjecture is that the fundamental objects in the universe
are neither particles, nor waves, but field patterns. In this regard, it seems likely that the associated space-time microstructure would have a length scale comparable to the Planck length (about $10^{-35}$ meters) and a time scale comparable to the Planck time (about $10^{-44}$ seconds).

Space-time microstructures are  composites whose microstructure varies not just with respect to space but also with respect to time. There exist two 
natural ways of changing the material properties in time, thus, leading to two types of space-time microstructure. Using the terminology in the book of
Lurie \cite{Lurie:2007:IMT} \textit{activated materials} (the subject matter of this article) are immovable with respect to the laboratory frame and the space-time microstructure is realized by an external mechanism that produces a time switching (either instantaneous or gradual) of the space pattern of the material in a pre-determined manner; in \textit{kinetic materials}, instead, the space-time microstructure is realized by an actual mechanical motion of the various parts of the composite system with respect to the laboratory frame. In other words, in activated materials the motion (without transfer of matter) is only in terms of the property pattern, whereas in kinetic materials the  space-time microstructure is achieved by moving fragments of the material assemblage. Clearly, standard composites can be considered as  space-time microstructures with a microstructure that depends only on space.
Activated materials are most easily achieved (see, for example, \cite{Louisell:1958:PAS}) by propagating a large amplitude wave (the "pump wave") into a non-linear homogeneous, or possibly inhomogeneous, medium. One could even propagate several large amplitude waves that then form
an interference pattern. On top of these large waves one superimposes waves that are sufficiently small that one can linearize the problem and treat them as if they are propagating in a medium with space-time variations given by the tangent moduli associated with the large amplitude waves. There are many other ways of creating space-time geometries too. For instance, one could consider a spatially periodic two-phase
geometry where one phase is a liquid crystal and then subject this material to an oscillating electric field that then causes a temporal modulation in the refractive index of the liquid crystal phase.

Space-time microstructures were studied as early as 1958 by Cullen \cite{Cullen:1958:TWP}, who noted that a transmission line, with modulated inductance that propagates along the line, could support a current wave that 
is periodic in time but grows exponentially along the transmission line (in space). This transmission line, with time varying modulated inductance, can be viewed as a "space-time laminate with continuously varying moduli" in which the moduli are just functions of $\Bn\cdot\Bx$, where $\Bx$ and $\Bn$ are two-dimensional vectors and we identify the components of $\Bx$, $x_1$ with space and $x_2$ with time.  If one interchanges the roles of time and space (switching $x_1$ and $x_2$) in the analysis of Cullen, one naturally gets a "space-time laminate" which supports a field that is exponentially
growing in time. Shortly afterwards, space-time laminates that interact with a field $\Bu=(V_1,V_2)$ consisting of a pair of potentials $V_1$ and $V_2$ were investigated by Tien \cite{Tien:1958:PAF} who found that they could transfer power between waves at different frequencies. Morgenthaler \cite{Mor:1958:VME}, while studying electromagnetic wave propagation in a homogeneous dielectric medium with time-dependent permittivity and permeability, derived a solution for the cases of a two-layered temporal laminate and for a temporal graded material. Subsequently, Fante \cite{Fan:1971:TEW} proposed an explicit solution for the electromagnetic problem of a multi-layered temporal laminate with layers having the same wave impedance, and a solution in closed form for the case of layers with different wave impedances.
Many more early references and an experimental validation can be found in the paper of Honey and Jones \cite{Honey:1960:WBU}.


Activated space-time laminates (where the moduli only depend on $\Bn\cdot\Bx$ in which $x_{d+1}$ is identified with time, and $d$ is the number of spatial
dimensions)
are of particular interest as they exhibit remarkable properties: indeed, by suitably controlling the design parameters it is possible to selectively screen large space-time domains from long wave disturbances \cite{Lurie:1997:EPS,Weekes:2001:NCW}, an effect that is impossible to realize with standard laminates. 
In particular, waves cannot travel in the forward direction when the microstructure is effectively moving faster than the speed of wave
propagation in an equivalent stationary medium. Contrary to what happens in other space-time microstructures, the total energy of two-component 
space-time laminates is preserved for low frequency waves: the energy pumped into the system to switch from one material to the other (say, from material 1 to material 2) is equal to the energy released at the subsequent switch (from material 2 to material 1) \cite{Lurie:2003:SNA}. For such low frequency waves one can replace the laminated microstructure with a homogeneous one with suitable effective properties. The determination of these effective
properties \cite{Lurie:1997:EPS,Lurie:1998:PEP,To:2009:HDL}
follows the standard procedure for static laminates (see, for example, Chapter 9 of \cite{Milton:2002:TOC} and references therein)
for which the homogenized parameters are easily derived through direct calculation. When the frequency is not low, one can regard the laminated structure
as a spatially periodic structure, independent of time, that is rotated in space-time. Thus one can
apply standard Bloch-Floquet theory, with a harmonic dependence on time replaced with a harmonic dependence on $\Bm\cdot\Bx$ where $\Bm$ is perpendicular to $\Bn$. An additional assumption, which permits a lot of explicit analysis, is to assume that the fluctuations in the moduli are sinusoidal,
only involving one Fourier component. Then the constitutive relation only couples together neighboring Fourier modes, and is easily solved
\cite{Cassedy:1963:DRT,Cassedy:1965:WGB,Chu:1972:WPD}. For high frequency waves the energy balance need not occur, and waves can grow exponentially in time 
\cite{Cassedy:1967:DRT}.

Besides space-time laminates, other space-time microstructures have been proposed in the literature. In particular, special attention has been drawn towards rectangular microstructures in one spatial dimension and time (e.g., \cite{Lurie:2006:WPE} and \cite{Lurie:2009:MAW}): the geometry is doubly-periodic in time and space, and the space period is of the same order of the time period. Specifically, the case of checkerboard geometries where the two constituent materials have the same wave impedance, so that there is not a reflected wave but only a transmitted wave, has been extensively investigated (e.g., \cite{Lurie:2006:WPE}). Very interestingly, within each space period, the disturbances converge towards a so-called ``limit cycle'' after a few time periods, if the parameters of the constituents are suitably chosen. As a group of characteristics converges to a limit cycle, the energy of the macroscopic system grows exponentially (see, e.g., \cite{Lurie:2016:EAW}). It is almost
like shocks are developing in a linear medium. (One wonders if a stochastic version of such linear shocks could be a mechanism for the collapse of the wave function in quantum mechanics. Note that at the Planck length scale it is questionable as to whether the concept of energy has any meaning). In particular, the spatial derivative of the disturbance increases every time the wave passes through a pure spatial interface, whereas its time derivative grows every time the wave passes through a pure time interface \cite{Lurie:2006:WPE}. Therefore, unlike space-time laminates at low frequencies, space-time checkerboard geometries accumulate energy independently of the frequency, and, clearly, this does not allow one to apply the standard homogenization techniques to determine the effective properties of the material. Nevertheless, the macroscopic behavior of such exponentially growing fields is probably described by
some coarse grained equations. (We also find exponentially growing fields, yet it seems likely that macroscopic features, such as the conical shape seen later
in \fig{num_logc100}, might be describable by some coarse grained equations.)
It is worth noting that another consequence of the exponentially accumulation of energy in macroscopic systems is the thermodynamic non-equilibrium of such materials which, on the contrary, seem to be thermodynamically open systems: only when analyzed together with the surrounding environment can they be considered as thermodynamically closed systems. We should point out that the crossing of pure temporal interfaces, in correspondence to which the properties of the material change instantaneously, leads to very interesting phenomena, such as the Doppler effect analyzed, for instance, by Rousseau et al. \cite{Rousseau:2011:ESD}.

The extension of the theory to space-time microstructures with properties varying in a two-dimensional space and in time reveals new aspects totally absent in the one-dimensional case. For instance, in the work of Sanguinet \cite{Sanguinet:2011:HEM}, the homogenization of an elastic laminate in plane strain (with both inertial and elastic properties varying in time and space) leads to two new additional forces, one of which is of the Coriolis-type. This force, arising from the dynamics and the plane strain hypothesis, vanishes when the model is restricted to the one-dimensional case. 

On the other hand, specific unconventional space-time microstructures have been designed to optimize certain properties by means of topology optimization. Following the pioneering papers of Maestre et al. \cite{Maestre:2007:STD} and Maestre and Pedregal \cite{Maestre:2009:STD}, where the optimization of the distributions of materials in one-dimensional and two-dimensional space and in time has been studied, Jensen \cite{Jensen:2009:STT} proposed an optimized
dynamic structure with time varying stiffness that prohibits wave propagation: it consists of a moving bandgap  with layers of stiff inclusions
moving with the propagating wave. These results have been also extended to the case of time varying mass density \cite{Jensen:2010:OST}.

As space-time microstructures can break time reversal invariance, nonreciprocal frequency conversion \cite{Yu:2009:COI,Lira:2012:EDN} and other nonreciprocal effects (see \cite{Taravati:2016:MDA} and references therein) can occur. Moreover, most remarkably, Fang, Yu, and Shan \cite{Fang:2012:REM} show that one can get effective magnetic fields for photons by dynamic modulation. Also Yuan, Shi, and Fan \cite{Yuan:2016:PGP},
following related ideas of Boada, Celi, Latorre and Lewenstein \cite{Boadi:2012:SGF}
and Celi, Massignan, Ruseckas, Goldman, Spielman, Juzeli\ifmmode \bar{u}\else \={u}\fi{}nas and M. Lewenstein \cite{Celi:2014:SGF},
find that it is useful to introduce an extra "synthetic frequency" dimension in the modeling of the behavior of arrays of resonators undergoing a time-harmonic refractive index modulation, and this has lead to interesting ideas such as
optical models of four-dimensional quantum Hall physics \cite{Ozawa:2016:SDI} and simulated Weyl points
\cite{Zhang:2016:PAI}.

Irrespective of the fascinating possible connection of the theory of field patterns to quantum mechanics, 
the theory of field patterns is intrinsically interesting and 
could prove relevant in the study of spatially periodic 
composites of hyperbolic materials. Hyperbolic materials are materials where the dielectric tensor has both positive and negative eigenvalues. They were studied in the context of anisotropic plasmas by Fisher and Gould \cite{Fisher:1969:RCF} back in 1969, albeit with an antisymmetric part added to the dielectric tensor. 
In any physical hyperbolic material these eigenvalues also have a small imaginary part that causes absorption of electromagnetic energy into heat.  The 
simplest way to obtain hyperbolic materials is simply to laminate an isotropic material with positive dielectric constant with a material with
negative dielectric constant at the frequency under consideration: candidates for materials with negative dielectric constant over a frequency range
include silver, gold, and silicon carbide as well as host of novel materials \cite{Naik:2013:APM}.  If one assumes that homogenization theory applies (and the conditions for this assumption to be valid warrant further investigation), then the effective
dielectric constant of the laminate is given by the arithmetic average in any direction parallel to the layers, and by the harmonic mean in the direction
orthogonal to the layers. The key point is that the harmonic averages and arithmetic averages are sometimes of opposite sign, giving rise to a
hyperbolic effective dielectric tensor. There are also naturally occurring hyperbolic
materials \cite{Korzeb:2015:CNH}. Hyperbolic materials have generated considerable interest as they can direct radiation along the "characteristic lines" in the hyperbolic medium. With the hyperbolic medium in a shell configuration, and the material oriented so an eigenvector of the dielectric tensor
points in the radial direction, sources near the  inner boundary of the shell can be spaced less than a wavelength apart, yet radiate along the characteristic lines to the outer surface of the shell where they can be greater than half a wavelength apart and thus detectable through conventional microscopic techniques. Thus this "hyperlens", proposed by Jacob, Alekseyev, and Narimanov \cite{Jacob:2006:OHF}
and by Salandrino and Engheta \cite{Salandrino:2006:FFS}, that was subsequently experimentally validated \cite{Rho:2010:SHT,Lu:2012:HMF}, 
allows one to resolve sources that are less than a wavelength apart. In the quasistatic regime, where the wavelength is much larger than the body
under consideration, the governing equations are 
\beq \Bd(\Bx)=\BGve(\Bx)\Be(\Bx),\quad{\rm where}\quad \Div\Bd=0,\quad \Be=-\Grad V, \eeq{0.1}
where $\Bd(\Bx)$, the electric displacement field, $\Be(\Bx)$, the electric field, $V(\Bx)$, the electric potential, and $\BGve(\Bx)$ the 
dielectric tensor, are all complex valued. Consider the simplest case where the material and fields are two-dimensional (or three-dimensional but
independent of one spatial coordinate), and the dielectric tensor is diagonal and hyperbolic,
\beq \BGve(\Bx)=\bpm \Ga(\Bx) & 0 \\ 0 & -\Gb(\Bx) \epm, \eeq{0.2}
in which $\Ga(\Bx)$ and $-\Gb(\Bx)$ are complex valued functions with non-negative imaginary part. Then \eq{0.1} reduces to 
\beq \frac{\Md}{\Md x_1}\left[\Ga(\Bx)\frac{\Md V}{\Md x_1}\right]=\frac{\Md}{\Md x_2}\left[\Gb(\Bx)\frac{\Md V}{\Md x_2}\right].
\eeq{0.3}
In regions where $\Ga(\Bx)$ and $\Gb(\Bx)$ are constant, real, and positive this is just the standard wave equation if one reinterprets
$x_2$ as a time coordinate, keeping $x_1$ as a spatial coordinate. So a spatially periodic hyperbolic material where the dielectric
tensor takes the form \eq{0.2} can be viewed in the limit where the loss goes to zero as a space-time microstructure. An interesting question
is whether in spatially periodic composites of hyperbolic materials one can get unusual homogenized equations in the limit where the imaginary
part of $\Ga(\Bx)$ and $\Gb(\Bx)$ tends to zero. One might expect that such unusual homogenized behavior could occur if the "cell-problem",
where one looks for periodic $\Bd(\Bx)$ and $\Be(\Bx)$ that solve \eq{0.1} with a prescribed non-zero average value of $\Be(\Bx)$,
has a nonunique solution when the loss is zero.

We should point out that besides the study of space-time microstructures, there is a large body of work on 
active control by using structures with moduli that vary in space and time (see, for instance, \cite{Krylov:1997:DEB}, \cite{Sorokin:2000:ACV}, and \cite{Sorokin:2004:AWV}). Also, interestingly, space-time boundaries have been found to be important for time reversal \cite{Fink:2016:LDT,Goussev:2016:LET}: a wave pulse hitting a space-time boundary divides into two wave pulses: one continues to spread outwards and the other converges to the position of the original source. This division of wave pulses is also apparent in the work of Lurie and coworkers \cite{Lurie:1997:EPS,Weekes:2001:NCW,Blekhman:2007:OVD,Lurie:2009:MAW}.

Subsequent to the initial phases of this work it was discovered by Alexander and Natalia Movchan that field patterns occur in space-time geometries that are as simple as a two-phase laminate with boundaries that are normal to the time axis (i.e, at a periodic set of times). The analysis of this will be explored elsewhere \cite{Movchan:2016:FPT}.

\section{Statement of the problem}
\setcounter{equation}{0}

To begin with, and to keep things simple, our first interest is in the two-dimensional conductivity equation
\beq \Bj(\Bx)=\BGs(\Bx)\Be(\Bx),\quad{\rm where}\quad \Div\Bj=0, \quad \Be=-\Grad V, \eeq{1.1}
with $\Bj(\Bx)$ the electric current, $\Be(\Bx)$ the electric field, $V(\Bx)$ the electric potential, and $\BGs(\Bx)$ the conductivity tensor. Here we focus only on the conductivity problem of an assemblage of two materials, 
that is, we study the case where $\BGs(\Bx)$ can take only two values:
\beq \BGs(\Bx)=\Gc(\Bx)\BGs_1+[1-\Gc(\Bx)]\BGs_2, \eeq{1.2}
where $\Gc(\Bx)$ is a characteristic function that takes the value 1 in the region of the material where $\BGs(\Bx)=\BGs_1$ and takes the value 0 in the region where $\BGs(\Bx)=\BGs_2$,
with the conductivities $\BGs_1$ and $\BGs_2$ taking the form
\beq \BGs_1=\bpm \Ga_1 & 0 \\ 0 & -\Gb_1 \epm,\quad
\BGs_2=\bpm \Ga_2 & 0 \\ 0 &  -\Gb_2 \epm, \eeq{1.3}
where the parameters $\Ga_i$ and $\Gb_i$, $i=1,2$ are, in general, real and positive. The reason why the conductivity coefficient in the $x_2$ direction is taken to be negative is merely for the sake of convenience. In fact, in such a way, it is easy to see that by combining equations \eq{1.1} and \eq{1.3}, the potential $V_i(\Bx)$ in phase $i$, $i=1,2$, satisfies the following wave equation
\beq \Ga_i\frac{\Md^2 V_i}{\Md x_1^2}= \Gb_i\frac{\Md^2 V_i}{\Md x_2^2}. \eeq{1.4}
Indeed, we can think of $x_1$ representing the space variable $x$, and $x_2$ as representing the time variable $t$. Consequently, the parameters $\Ga_i$, $i=1,2$, are nothing but the usual conductivity coefficients in space, whereas the $\Gb_i$, $i=1,2$, have to be interpreted as the conductivity coefficients in time. In other words, we are considering a one-dimensional distribution of the two materials and we suppose that such a configuration varies not just with respect to the spatial coordinate $x$ but also with respect to time, thus giving rise to a dynamic material.  

It is well-known that the local solution in phase $i$ of equation \eq{1.4}, deduced via d'Alembert method, is simply given by 
\beq V_i(x,t)=V^+_i(x-c_it)+V^-_i(x+c_it) \eeq{1.5}
where $V^+_i(x-c_it)$ is the wave moving upwards to the right in a space-time diagram, $V^-_i(x+c_it)$ is the wave moving upwards to the left in a space-time diagram, and
\beq c_i=\sqrt{\Ga_i/\Gb_i} \eeq{1.6}
is the wave space. Note that associated with $V^+_i(x-c_it)$ and $V^-_i(x+c_it)$ are currents 
\beqa \Bj^+_i=-\BGs_i\Grad V^+_i(x-c_it)=\sqrt{\Ga_i\Gb_i}\bpm -c_i \\ -1 \epm {V^+_i}'(x-c_it), \nonum
\Bj^-_i=-\BGs_i\Grad V^-_i(x+c_it)=\sqrt{\Ga_i\Gb_i}\bpm -c_i \\ 1 \epm {V^-_i}'(x+c_it),
\eeqa{1.7}
flowing along the characteristic lines, where ${V^+_i}'(s)$ and ${V^-_i}'(s)$ denote the derivatives
of ${V^+_i}(s)$ and ${V^-_i}(s)$. It is clear that the currents do not interact and the potentials do not interact and, therefore, we can, in a sense, think of the medium as composed of two parallel independent sets of wires aligned with the characteristic directions (see \fig{0}).  

Clearly, the explicit expression of the d'Alembert solution \eq{1.5}, and therefore of \eq{1.7}, depends on the initial conditions chosen (see Section \ref{Initial_cond_sec}). For what concerns the boundary conditions we assume here that the medium is infinite with respect to $x$. (Later, in the numerical section, we will assume periodic boundary conditions
in $x$.)

\begin{figure}[!ht]
\centering
\includegraphics[width=0.7\textwidth]{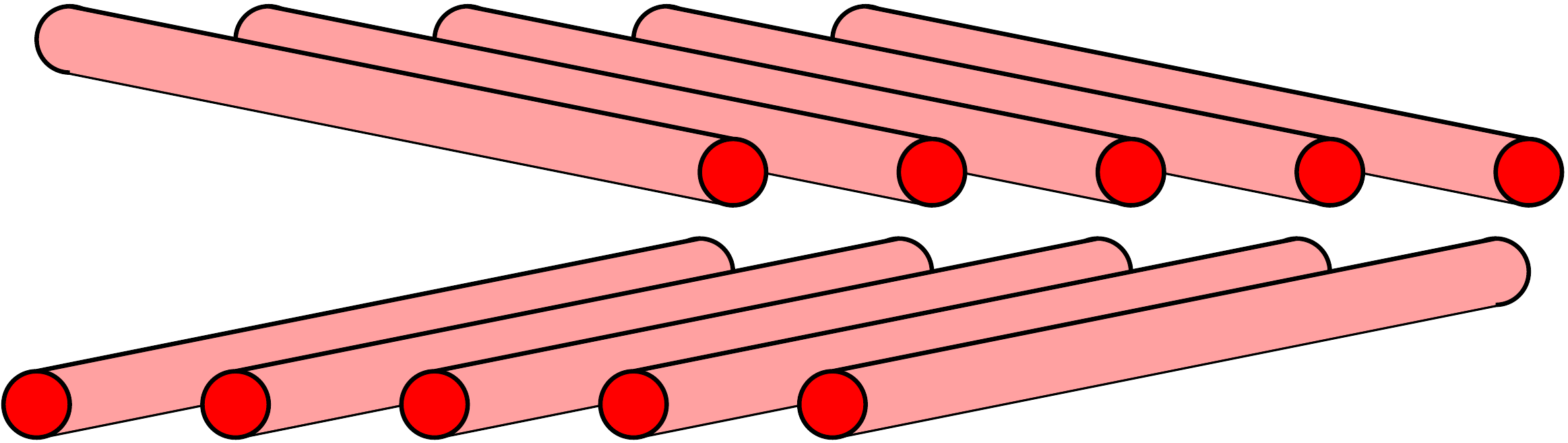}
\caption{A hyperbolic medium can be thought as if it is composed of sets of unconnected wires, each set in the direction of the characteristics.}
\labfig{0}
 \end{figure}

\subsection{Transmission conditions at the interfaces}

First of all, we recall that, in principle, the space-time distribution of the two materials is arbitrary, provided that the existence and uniqueness of the solution $V(x,t)$ is ensured. This places a constraint on the shape of the interfaces: each interface has to be such that there are always two incoming and two outgoing characteristics (see, for instance, \cite{To:2009:HDL}). To attain such a requirement, it is sufficient that the slope $w$ of each interface in a space-time diagram fulfills the following relation \cite{Lurie:1997:EPS}:
\beq (w^2-c_1^2)(w^2-c_2^2)\geq 0\,. \eeq{condition_w}
In case of a pure spatial interface, that is, in case the interface is a vertical line in a space-time diagram, we have $w=0$ and the above condition is trivially satisfied. Similarly, in case of a pure temporal interface, corresponding to a horizontal line in a space-time diagram, we have that $w\to\infty$ and once again condition \eq{condition_w} is trivially fulfilled.

For what concerns the transmission conditions for the potential across the interfaces, we require that the potential  be continuous across each interface and such that the continuity of the current flux is preserved, i.e.,
\beqa V_1=V_2
\label{continuity_V}\\ \Bn\cdot\BGs_1\Grad V_1=\Bn\cdot\BGs_2\Grad V_2\,,\eeqa{Continuity_flux}
with $\Bn$ being the normal vector to the interface.

\subsection{Initial conditions}\label{Initial_cond_sec}

Let us start by considering initial conditions of the type
\beq V(x,0)=H(x-a), \quad j_2(x,0)=\Gd(x-a)j_0, \eeq{1.7a}
and see how the corresponding disturbance propagates. Here $H(y)$ is the Heaviside function,
\beqa H(y) & = & 0\quad{\rm if}~ y<0, \nonum
 & = & 1\quad{\rm if}~ y>0, 
\eeqa{1.8}
and $\Gd(y)$ is the Dirac delta function. Thus, we are injecting a total current flux $j_0$ at time $t=0$ concentrated at $x=a$. In general, as illustrated in \fig{0a},
this generates a cascade of current lines
branching as $t$ increases that is difficult to analyze. If the process is ergodic, then one can probably understand the dynamics in an ensemble averaged sense, similarly to what is done in statistical physics. 

\begin{figure}[!ht]
\centering
\includegraphics[width=0.7\textwidth]{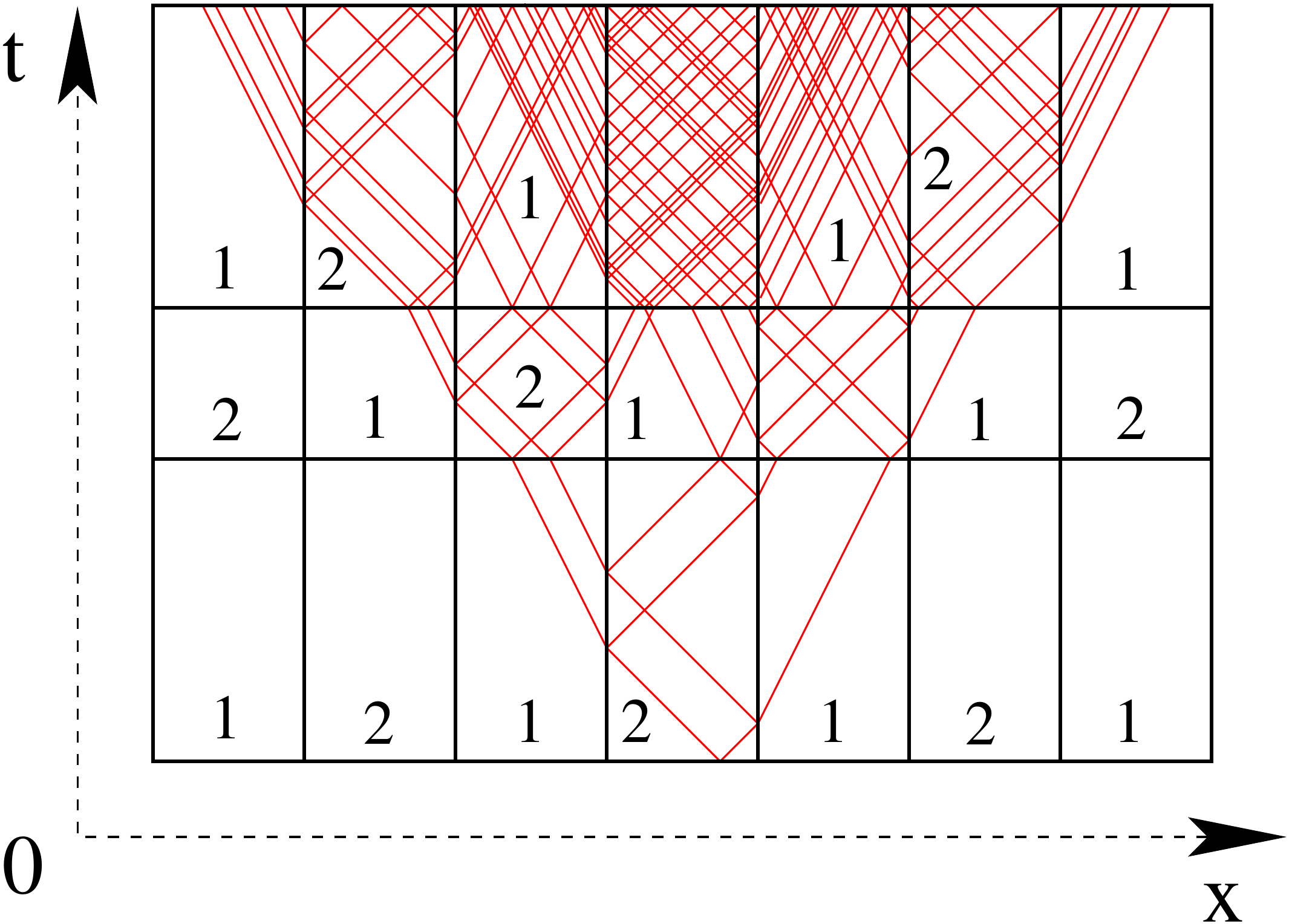}
\caption{A typical dynamical process, showing a complicated cascade of current lines.}
\labfig{0a}
 \end{figure}

\subsection{Field Patterns}
What is remarkable is that there are special space-time geometries, as illustrated in \fig{1} and \fig{2}, where the dynamics is especially simple (note that the restriction \eq{condition_w} is trivially satisfied by both geometries having only horizontal and vertical interfaces). 
In particular, what causes the orderly pattern of characteristics in \fig{1} and in \fig{2} is the special relation between the characteristic lines and the geometry of the microstructure, as displayed in \fig{1.5}. From the configuration of the characteristic lines it is clear that, denoted with $x_0$ and $t_0$ the spatial and time dimensions of the unit cell of each space-time microstructure, their ratio reads
\beq x_0/t_0=\frac{c_1(c_1+2c_2)}{c_1+c_2}, \eeq{1.8a}
for the geometry shown in \fig{1}, and
\beq x_0/t_0=\frac{c_1(c_1+3c_2)}{2c_1+c_2}, \eeq{1.8ab}
for the microstructure presented in \fig{2}. We will see that the dimensions of the unit cell of the space-time microstructure could be different from those of the unit cell of the network dynamics (for instance, for the microstructure in \fig{1}, the former are $x_0$ and $t_0$ and the latter are $2x_0$ and $t_0$). 

Note that by fixing the microgeometry we fix the volume fraction $f$ of the space-time inclusions which, for the microstructure in \fig{1}, is equal to $f=c_1c_2/[(c_1+2c_2)(c_1+c_2)]$, while for the geometry in \fig{2} is equal to $f=c_1c_2/[(c_1+3c_2)(c_1+c_2)]$.

\begin{figure}[!ht]
\centering
\includegraphics[width=0.7\textwidth]{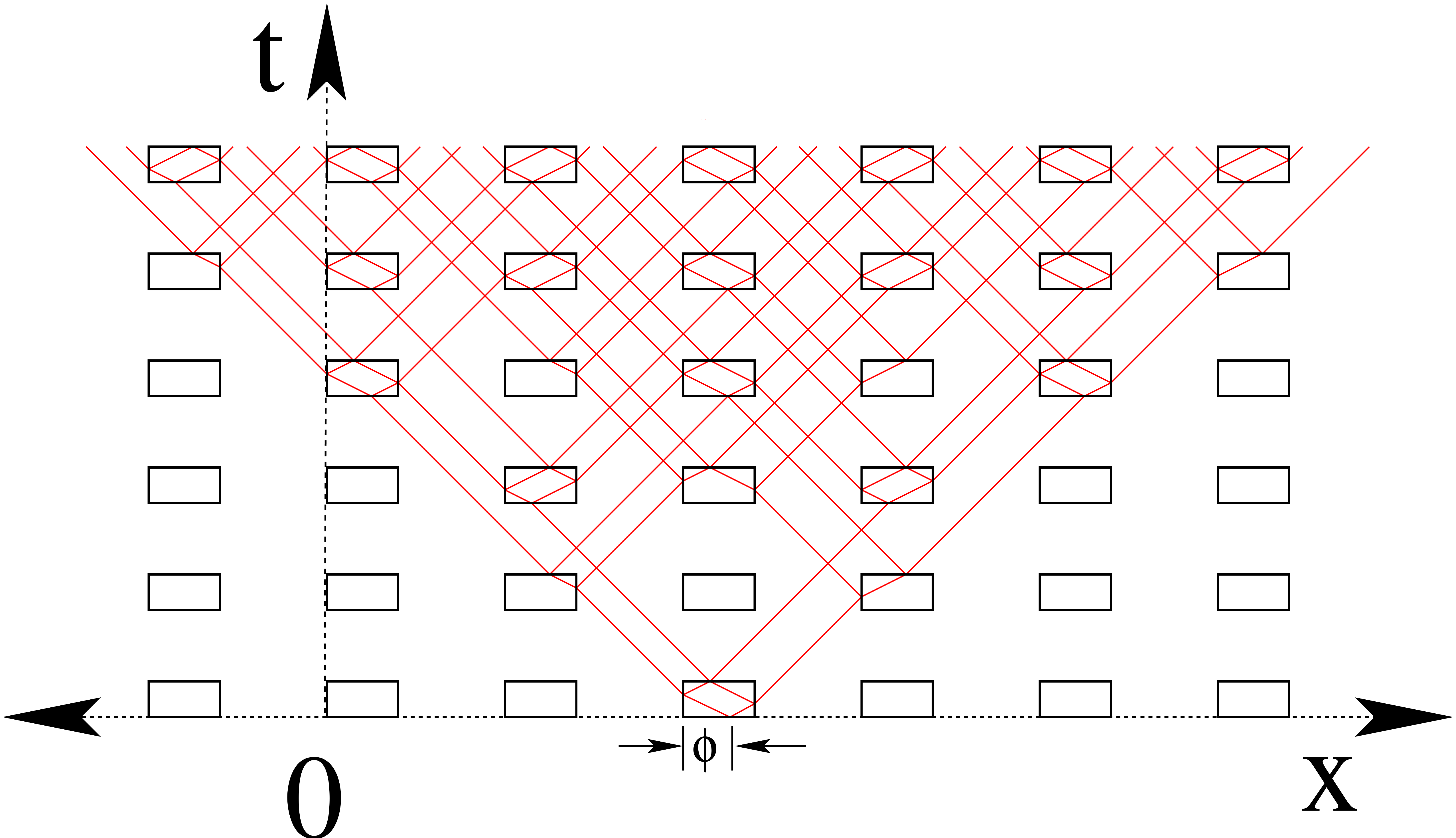}
\caption{If the space-time geometry is suitably chosen, then the dynamics could be very simple and exhibit interesting features. For instance, here we show the case of a dynamic material composed of a matrix of material 1 with space-time rectangular inclusions of material 2. The distance between the inclusions is set in such a way that this periodic microstructure gives rise to a special orderly dynamics. The parameter $\phi$ is called the index parameter and it 
paramaterizes the field pattern.}
\labfig{1}
\end{figure}
\begin{figure}[!ht]
\centering
\includegraphics[width=0.7\textwidth]{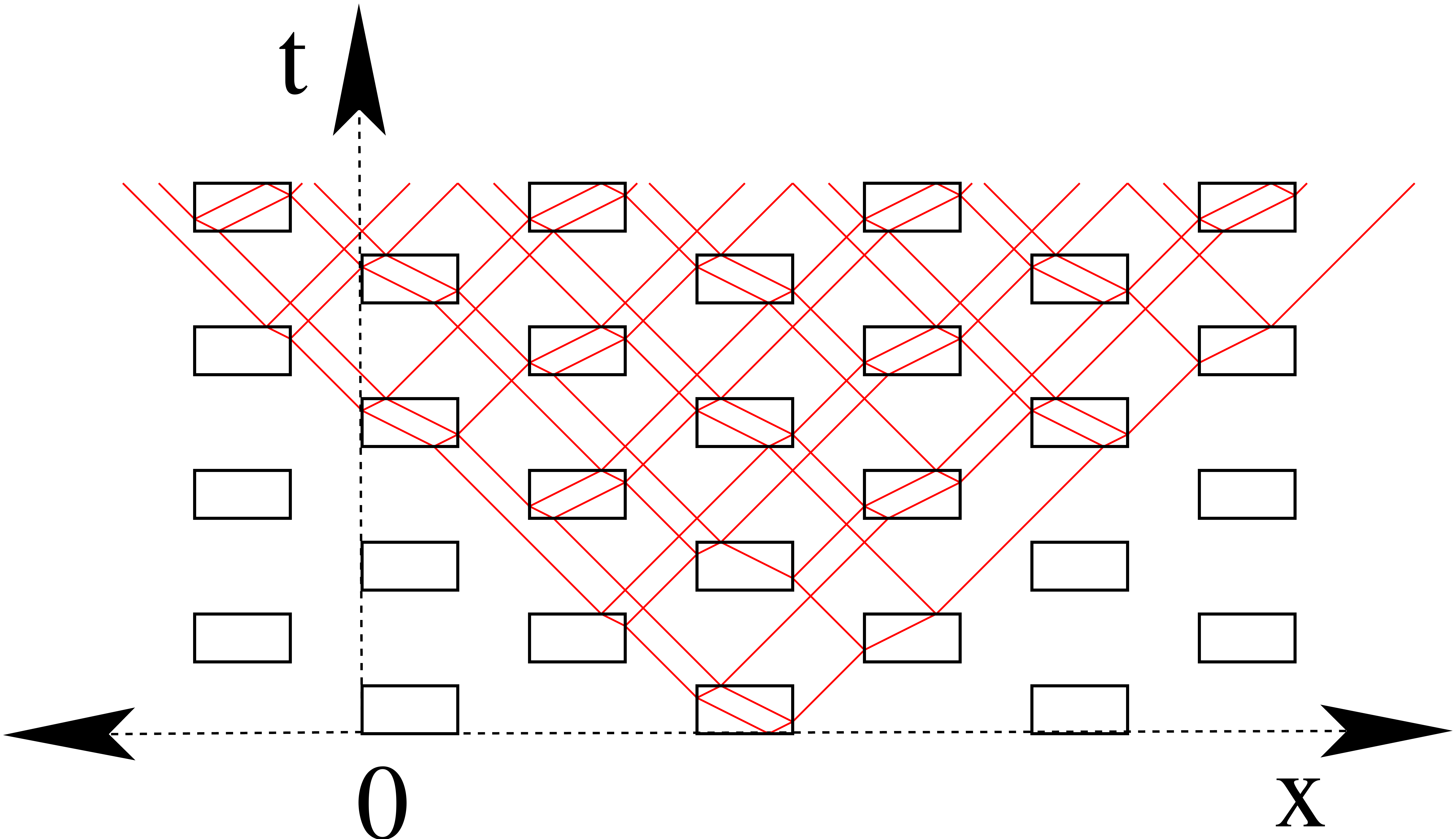}
\caption{Also in the case where the space-time rectangular inclusions are staggered (and not aligned as in the case shown in \fig{1}), the dynamics is very simple.}
\labfig{2}
 \end{figure} 
\begin{figure}[!ht]
\centering
\includegraphics[width=0.8\textwidth]{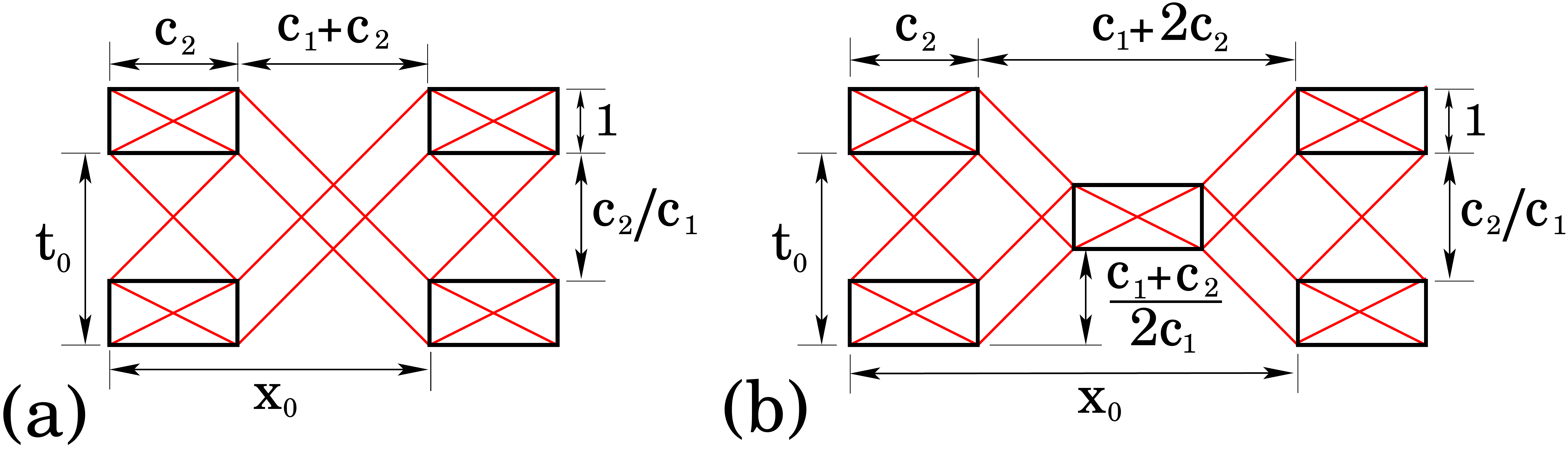}
\caption{The simplicity of the dynamics related to the space-time microstructures with aligned (\fig{1}) and staggered (\fig{2}) rectangular inclusions results from the specific geometrical structure of the two materials, as can be 
seen here, where the red lines denote characteristic lines.}
\labfig{1.5}
 \end{figure}

\section{Collections of field patterns as multidimensional or\\multicomponent objects} 
\setcounter{equation}{0}

Let us begin by focusing on \fig{1}, where the origin of the reference system $x=t=0$ corresponds to the bottom left corner of an inclusion. The periodic pattern of characteristic lines that arises after a sufficiently large amount of time is totally determined by the sole index parameter $\phi$, which is the distance between the bottom left corner of a specific inclusion and the intersection point between the base of that inclusion and the characteristic lines (see \fig{1}). Note that for all the other inclusions struck by characteristic lines, such a distance is either $\phi$ or $c_2-\phi$, where $c_2$ is the length of the base of the inclusion, if its height is set equal to one time unit. This is due to the fact that, for this space-time microstructure, the unit cell of the network dynamics (see also \fig{3} in the next Section) has length $2x_0$ and it contains two inclusions. Here, for simplicity, if we count the columns of inclusions starting from the right of the $t$-axis, those in an odd position will be parametrized by $\phi$, whereas those in an even position by $c_2-\phi$.

\begin{figure}[!ht]
		\begin{subfigure}{.5\textwidth}
		\flushleft
	    \includegraphics[width=0.94\textwidth]{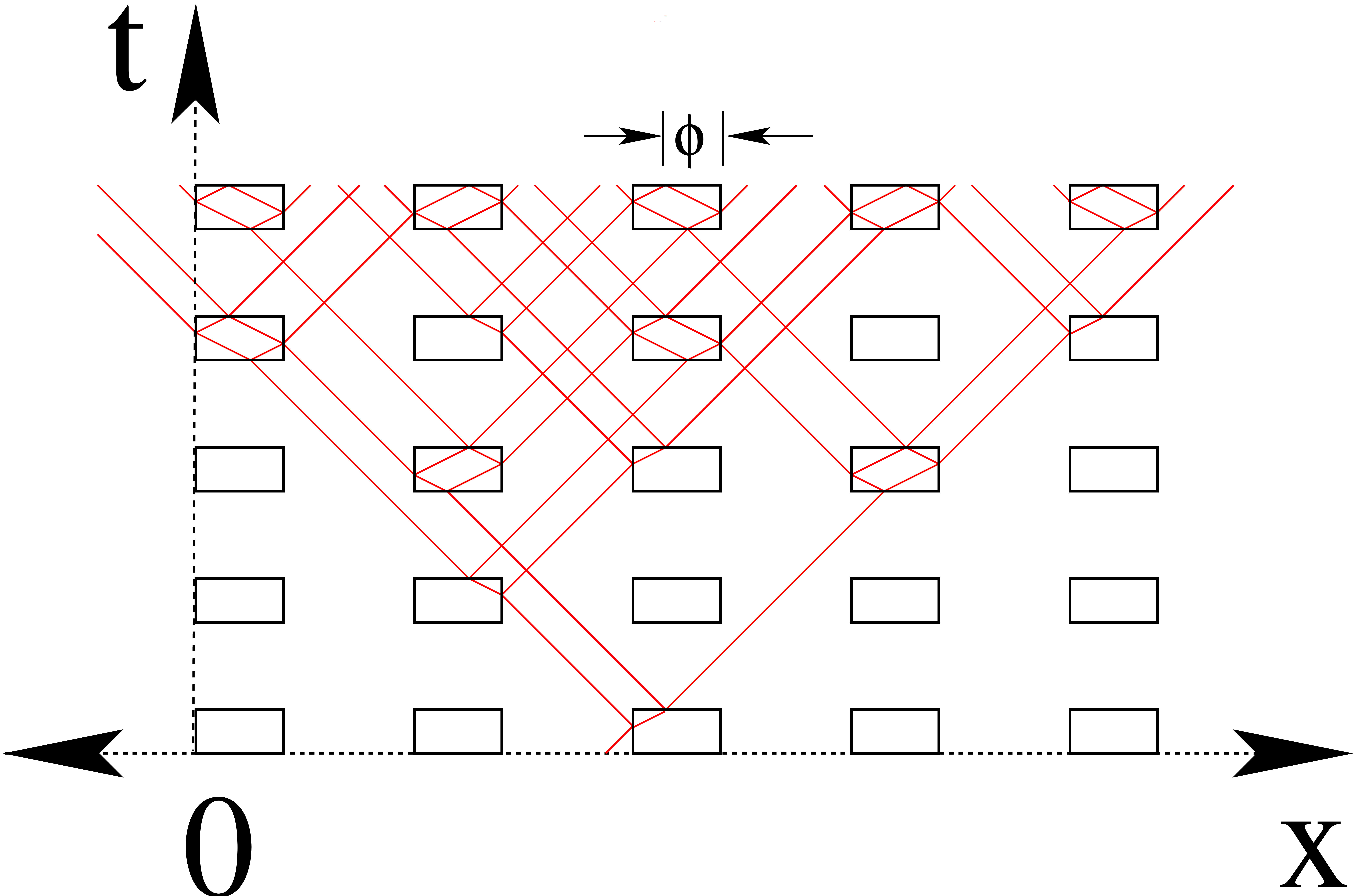}
	    \caption{}
	    \label{matrixfig}
	    \end{subfigure}%
        \begin{subfigure}{.5\textwidth}
        \flushright
        \includegraphics[width=\textwidth]{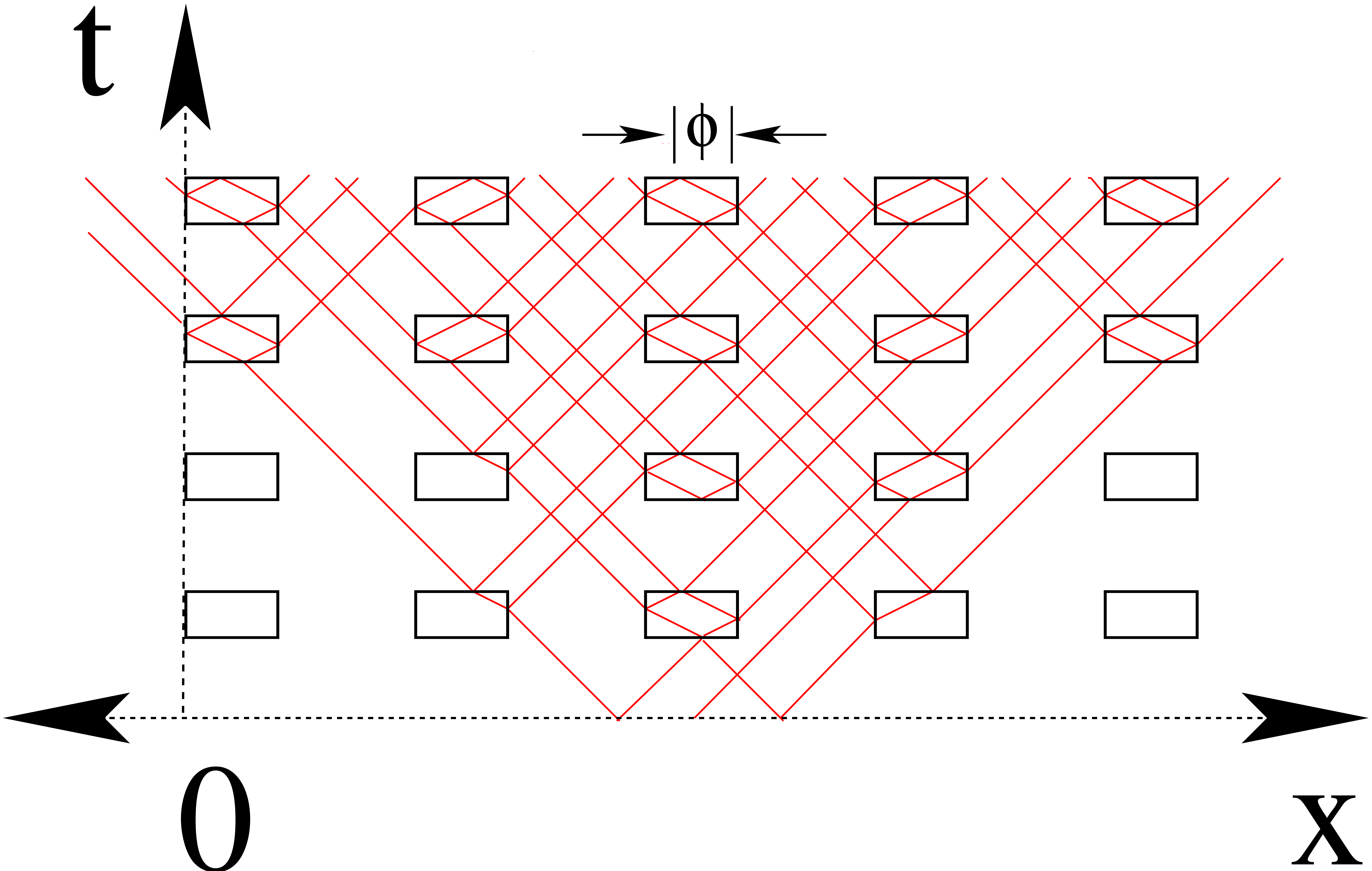}
        \caption{}
        \label{3pointsfig}
        \end{subfigure}%
    \caption{The field pattern of \fig{1}, indexed by the parameter $\phi$, can be obtained, for a sufficient large amount of time, also by considering 
initial conditions different from \eq{1.17}. In Figure \ref{matrixfig} the reference system is the same as in \fig{1} but the point of injection of 
current flux now belongs to the matrix, whereas in Figure \ref{3pointsfig} the $\phi$-field pattern is recovered by injecting current flux at three 
points in a different reference system. Note that these different launching conditions give rise to different current distributions within the field pattern.}
    \labfig{equivalent_field_patterns}     	
\end{figure}

Notice that the index parameter $\phi$ determines in a unique way the periodic pattern of characteristic lines that arises after a sufficiently large amount of time, as it is not related to the specific initial conditions chosen: the $\phi$-field pattern in \fig{1} can indeed be obtained in different ways and each way will have a different current distribution. For instance, in Figure \ref{matrixfig} the same $\phi$-field pattern of \fig{1} is launched by injecting current at a point $(a,0)$ belonging to the matrix, whereas in Figure \ref{3pointsfig} it is launched by injecting current at three points at a different time (note that in this figure the reference system is different from the one in \fig{1}). We will see in Section \ref{Numerical_result_alig} that some patterns parametrized by the same parameter $\phi$ blow up in time, while others have a wave-like behavior which does not blow up. 

If the initial conditions are suitably chosen, more than one field pattern can be launched. This gives rise to a collection of field patterns, each one parametrized by its index parameter $\phi_i$, with $i=1,2,\dots,m$, $m$ being the number of field patterns launched. For instance, pick two parameters $\Ge_1$ and $\Ge_2$ lying between $0$ and $1$, and choose real amplitudes
$s_1$, $s_2$, $j_1$, and $j_2$. Then, the initial conditions
\beqa\nonumber V(x,0)=s_1H(x-2x_0(l+\Ge_1))+s_2H(x-2x_0(n+\Ge_2)), \\ j_2(x,0)=\Gd(x-2x_0(l+\Ge_1))j_1+\Gd(x-2x_0(n+\Ge_2))j_2, \eeqa{1.9a}
where $l$ and $n$ are integers, can launch anywhere from one to four field patterns in the space-time microstructure of \fig{1}, depending on the value of the parameters $\Ge_1$ and $\Ge_2$. Suppose, first, that the points where the current is injected, $(x,t)=(2x_0(l+\Ge_1),0)$ and $(x,t)=(2x_0(n+\Ge_2),0)$, lie at the base of inclusions. If they both belong to inclusions in odd columns, we have $\Ge_i<c_2/(2x_0)=c_2/(2(c_1+2c_2))$, whereas if they both belong to an inclusion in an even column we have $1/2<\Ge_i<1/2+c_2/(2x_0)=1/2+c_2/(2(c_1+2c_2))$. In both cases, two field patterns can be launched
and the associated index parameters $\phi_1$ and $\phi_2$ are easily identified as $\phi_i=2\Ge_ix_0=2\Ge_i(c_1+2c_2)$, in the first case, and $c_2-\phi_i=2\Ge_ix_0-x_0=(2\Ge_i-1)(c_1+2c_2)$, in the second case. Note that if $\phi_1=\phi_2$ then only one field pattern is generated. Finally, if $c_2/(2(c_1+2c_2))<\Ge_i<1/2$ or $\Ge_i>1/2+c_2/(2(c_1+2c_2))$, then the points $(x,t)=(2x_0(l+\Ge_1),0)$ and 
$(x,t)=(2x_0(l+\Ge_1),0)$ and $(x,t)=(2x_0(n+\Ge_2),0)$ belong to the matrix and there could be up to
four associated field patterns: the index parameters are found by tracing the relevant characteristic or characteristics that originate from the source at $t=0$ until they strike the base of an inclusion.

Collections of field patterns are especially interesting because of their multidimensional nature. They are similar in some
respects to non-interacting particles in quantum mechanics. The idea is that one can split the space-time diagram into an infinite number of disjoint ``conducting networks'' that do not interact. Each field pattern in a collection then lives on
one of these conducting networks.

Consider a collection of $m$ field patterns, parameterized
by the index parameters $\phi_1$, $\phi_2$, $\ldots$, $\phi_m$. The potential $V_i(x,t)$ associated with the field pattern with index $\phi_i$ 
satisfies the wave equation and, by the superposition principle and the linearity of the problem, also the total potential
\beq V(x,t)=\sum_{i=1}^mV_i(x,t), \eeq{1.9aa}
satisfies the wave-equation. Although 
the dynamics of $V(x,t)$ is governed by the wave equation, it is far simpler to track the dynamics of 
$V_i(x,t)$ for each field pattern, $i=1,2,\ldots,m$, since the field patterns do not interact. We can think of $V_i(x,t)$ as living in its separate dimension,  $x_i$, and rather than considering the potential $V(x,t)$ one can consider the
multidimensional potential
\beq V(x_1,x_2,\ldots,x_m)=\sum_{i=1}^mV_i(x_i,t). \eeq{1.9ab}
This multidimensional potential is more suitable for capturing the dynamics than the underlying potential $V(x,t)$.
It is expected that one will be able, in appropriate conditions, to homogenize the behavior of field patterns
so that $V_i(x_i,t)$ is replaced by some suitably defined "coarse grained" potential $\underline{V}_i(x_i,t)$ 
that captures the macroscopic modulation of $V_i(x_i,t)$ and that satisfies an appropriate homogenized equation
(still to be found). While one could speak about the dynamics of the multidimensional coarse grained potential
\beq  \underline{V}(x_1,x_2,\ldots,x_m)=\sum_{i=1}^m\underline{V}_i(x_i,t), \eeq{1.9ac}
there would be no suitable equation to describe the dynamics of the coarse grained potential
\beq \underline{V}(x,t)=\sum_{i=1}^m\underline{V}_i(x,t), \eeq{1.9acd}
since too much information is lost in taking this last sum.

An alternative to introducing these extra dimensions is to introduce a vector potential $\BV(x,t)$ with
$m$ components $V_i(x,t)$. With coarse grained potentials $\underline{V}_i(x,t)$ the macroscopic behavior
of the collection of noninteracting field patterns is then governed by the vector potential $\underline{\BV}(x,t)$
having these potentials as components. If we think of these field patterns as living on separate "conducting networks"
then, in the presence of interactions (caused by small nonlinearities or a small imaginary part of $\BGs(x,t)$)
the work of  Khruslov \cite{Khruslov:1978:ABS} and Briane \cite{Briane:1998:HSW,Briane:1998:HTR}
(see also \cite{Boadi:2012:SGF} where related ideas are embodied in a quantum mechanical context)
suggests that the homogenized equations may simply couple these field components. If this turns out to be the case
then the full multivariate potential $\underline{V}(x_1,x_2,\ldots,x_m)$ may not be needed.

\section{Solving the cell problem}\label{Solving_cell_problem_section}
\setcounter{equation}{0}

In this Section we aim at making steps towards determining the effective properties of the material having the space-time microstructure shown in \fig{1}. The results concerning the microgeometry in \fig{2} are given in the supplementary material, as they are very similar to those presented here. We will just be concentrating on the case where the fields $\Be(\Bx)$ and $\Bj(\Bx)$ have the same periodicity
as the field pattern, i.e., their cell of periodicity contains two inclusions.  

As already noticed in the previous section, the special geometry in \fig{1} is such that the characteristic lines follows a certain periodic pattern. Let us consider, then, the unit cell for the network dynamics (see \fig{3}) (note that the size of such a unit cell is twice the size of the unit cell of the space-time microstructure, that is, $2x_0$ and $t_0$) and let us determine the solution of the problem on the unit cell in the case the initial conditions are
\beq V_i(x,0)=\frac{j_0}{c_i\,\Gb_i}H(x),\quad j_2(x,0)=j_0\Gd(x).\eeq{Initial_cond}
Consequently, the potentials $V^+_i(x,t)$ and $V^-_i(x,t)$ in phase $i$ take the following expressions:
\beq V^+_i(x,t)=a_i^+[1-H(x-c_it)],\quad V^-_i(x,t)=a_i^-H(x+c_it) \eeq{1.9} 
with $a_i^+=-{j_0}/(c_i\Gb_i)=-a_i^-$. The associated currents, according to \eq{1.7}, are
\beqa  \Bj^+_i=a_i^+\sqrt{\Ga_i\Gb_i}\bpm c_i \\ 1 \epm \Gd(x-c_it)\equiv \frac{a_i^+}{\Gg_i}\frac{1}{\sqrt{1+c_i^2}}\bpm c_i \\ 1 \epm \Gd(x-c_it), \nonum
\Bj^-_i=a_i^-\sqrt{\Ga_i\Gb_i}\bpm -c_i \\ 1 \epm \Gd(x+c_it)\equiv \frac{a_i^-}{\Gg_i}\frac{1}{\sqrt{1+c_i^2}}\bpm -c_i \\ 1 \epm \Gd(x-c_it),
\eeqa{1.10}
where 
\beq \Gg_i=\frac{1}{\sqrt{\Ga_i(\Ga_i+\Gb_i)}}. \eeq{1.11}
These currents flowing along the characteristics have magnitudes $j^+_i=a_i^+\Gg_i$ and $j^-_i=a_i^-\Gg_i$ in the direction of positive time. Bearing this remark in mind, since the jumps in the potential
across the characteristics in the direction of positive time are of magnitude $a^+_i$ and $a_i^-$ (see \eq{1.9}), then we can conclude that the constants $\Gg_1$ and $\Gg_2$ relate the potential jumps across the characteristic
lines to the current flowing through them.

We highlight the fact that the case analyzed in detail in this section is of particular interest as the solution of such a problem can be used as the starting point to build the solutions corresponding to the choice of other initial conditions, as also shown in Section \ref{Linear_potential}, where, instead of a prescribed jump in potential at $t=0$, we consider the case of a linear potential.

\begin{figure}[!ht]
\centering
\includegraphics[width=0.7\textwidth]{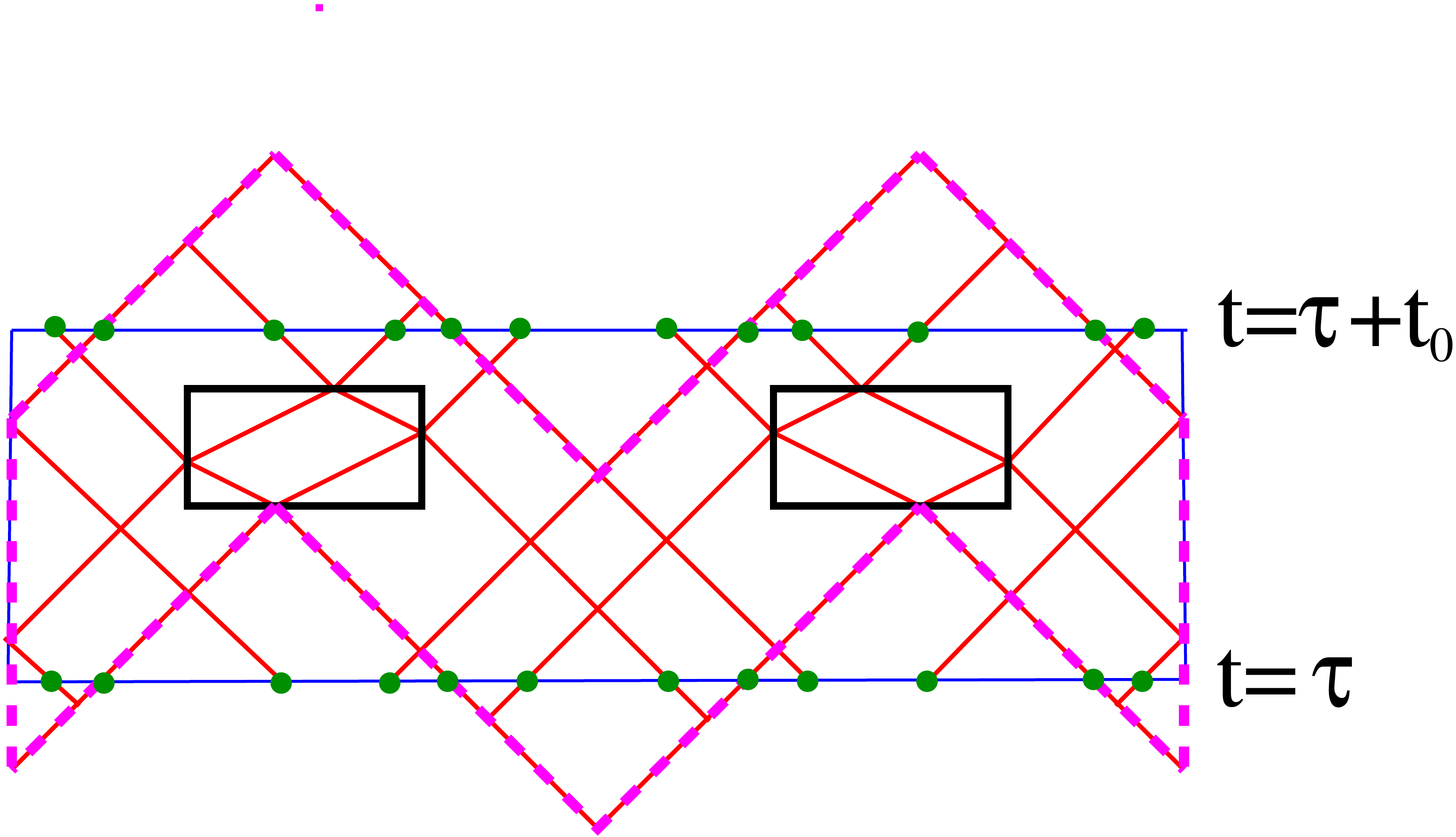}
\caption{Outlined by the purple dashed lines is the unit cell of periodicity for the dynamics of the microgeometry shown in \fig{1} in the case of a periodic field pattern.
The red lines denote characteristic lines. The current flows along such lines and, across them, there is a jump in potential. Note that the dynamics 
splits into dynamics that are symmetric and antisymmetric with respect to reflection about the vertical centerline of the cell. We call these the symmetric dynamics and the 
antisymmetric dynamics. If the field pattern is not periodic then we can still consider a periodic array of the characteristic lines, and the period cell could be taken to be that
outlined in blue. Current injected at time $t=\tau$ at any of the 12 points marked by the green dots at the bottom of the unit cell can contribute towards the excitation of the same field pattern. This current
exits the unit cell at time $t=\tau+t_0$ at any of the 12 points  marked by the green dots at the top of the unit cell, and may flow to the exit points of adjoining unit cells as well.}
\labfig{3}
\end{figure}

\begin{figure}[!ht]
\centering
\includegraphics[width=0.7\textwidth]{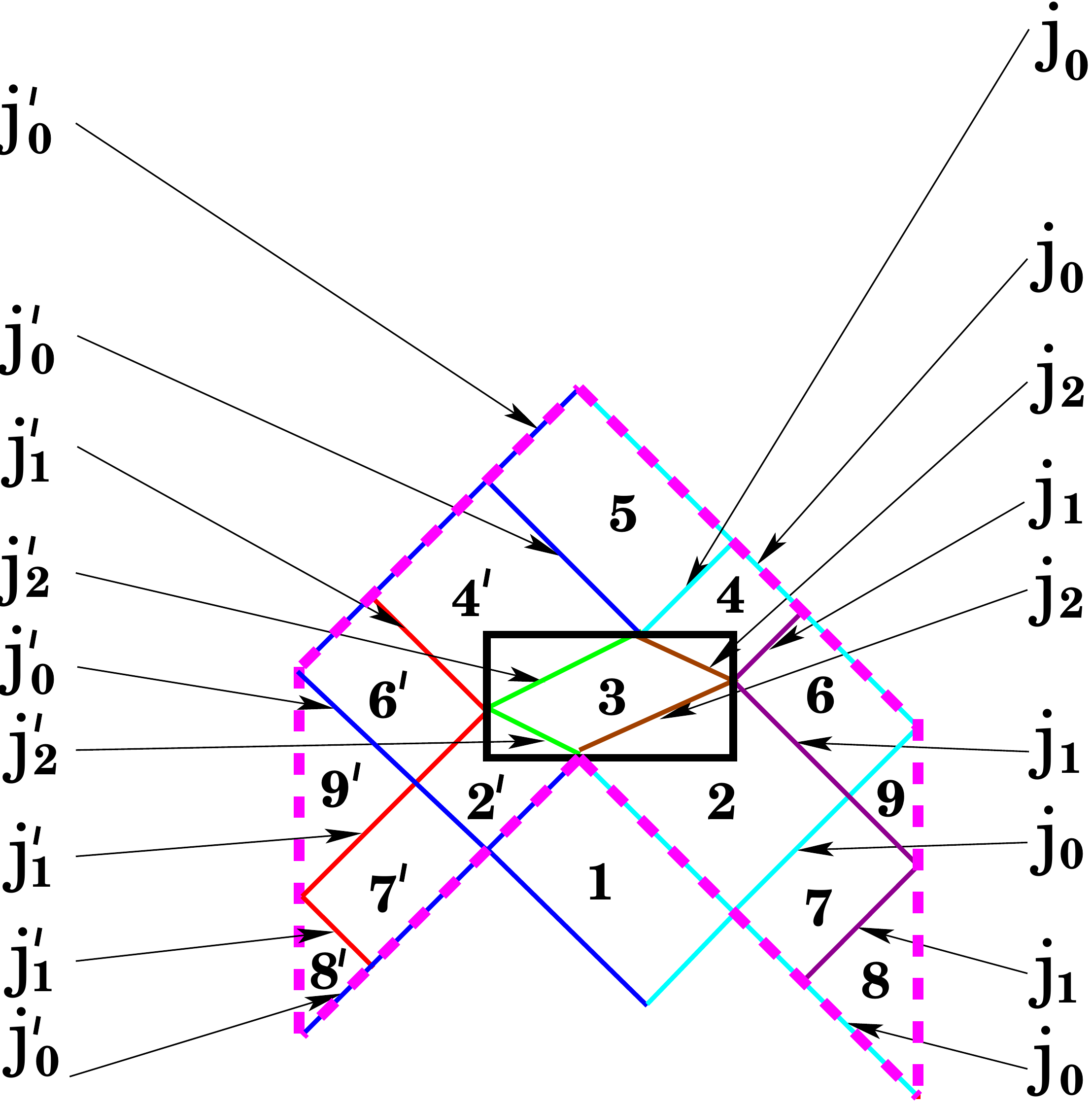}
\caption{The current flows in one half of the unit cell for symmetric dynamics. Here different colors are used to distinguish lines that have different currents flowing along them.
Periodicity and symmetry ensure that the currents flow as described in this figure.}
\labfig{4}
\end{figure}

The symmetry of the unit cell with respect to the vertical centerline suggests the idea of splitting the dynamics represented in \fig{3} into dynamics that are symmetric and dynamics that are antisymmetric: every type of dynamics can be recovered by suitably choosing the linear combination between these two elementary cases. 

\subsection{Symmetric dynamics}\label{Symm_section}

To solve for the currents and potentials associated with the symmetric dynamics with periodic currents (and hence periodic electric fields), we suppose all the currents
are flowing in the direction of positive time, and have the magnitudes $j_0$, $j_1$, $j_2$ and $j_0'$, $j_1'$, $j_2'$ as indicated on the lines in \fig{4}. Periodicity
and symmetry ensure the current flow has this structure. By conservation of current at the interfaces between the phases (see equation \eq{Continuity_flux}), we require that
\beq j_0+j_0'=j_2+j_2'. \eeq{1.12} 
Notice that, since the potential is piecewise constant, the condition of continuity given by \eqref{continuity_V} is trivially satisfied.

With reference to \fig{4}, we let $V_i$, $i=1,2,\ldots, 9$ denote the potentials in each of the regions $1,2,\ldots, 9$ and we let $V_i'$, $i=2,4,6,7,8,9$ denote the potentials in each of the regions
$2',4',6',7',8',9'$. For simplicity, we set $V_1=0$. Then, from the aforementioned rule that the potential jump across a characteristic line is $\Gg_i$ times the current flowing through it, we obtain
successively
\beq
\begin{aligned}
&V_2=\Gg_1j_0, \quad V'_2=\Gg_1j_0',\quad V_3=\Gg_1j_0+\Gg_2j_2=\Gg_1j_0'+\Gg_2j_2', \quad V_6=\Gg_1(j_0+j_1),\quad V_6'=\Gg_1(j_0'+j_1'), \\
&V_4 = \Gg_1j_0+2\Gg_2j_2=\Gg_1(j_0+2j_1),\quad V_4'=\Gg_1j_0'+2\Gg_2j_2'=\Gg_1(j_0'+2j_1'),
\quad V_7=0,\quad V_7'=0,\\& V_5=2\Gg_1j_0+2\Gg_2j_2=2\Gg_1j_0'+2\Gg_2j_2',\quad
V_8 = -\Gg_1j_1, \quad V_8'=-\Gg_1j_1', \quad V_9=\Gg_1j_1,\quad  V_9'=\Gg_1j_1'.
\end{aligned}
\eeq{1.13}
Thus, in addition to \eq{1.12}, we are left with the 3 constraints
\beq \Gg_1j_0+\Gg_2j_2=\Gg_1j_0'+\Gg_2j_2', \quad \Gg_2j_2=\Gg_1j_1, \quad \Gg_2j_2'=\Gg_1j_1' , \eeq{1.14}
which incidentally imply that $j_0+j_1=j_0'+j_1'$, as can be seen by multiplying the first formula in \eq{1.14} by $\Gg_1$.
So all the currents can be expressed in terms of only two independent currents, say $j_0$ and $j_0'$. Then,
\beqa  j_2& = &\frac{1}{2}(j_0+j_0')-\frac{\Gg_1}{2\Gg_2}(j_0-j_0'), \quad j_1=\Gg_2j_2/\Gg_1, \nonum
      j_2'& = &\frac{1}{2}(j_0+j_0')+\frac{\Gg_1}{2\Gg_2}(j_0-j_0'), \quad j_1'=\Gg_2j_2'/\Gg_1.
\eeqa{1.15}

The advantage of studying the symmetric dynamics in \fig{4}, rather than the general dynamics in \fig{3}, lies in the fact that the symmetry with respect to the centerline of the unit cell leads to average current fields and average electric fields that have non-zero components only in the
time direction. In particular, the average current field is easily worked out from the flux of current into the lower boundary of the cell of \fig{4}, 
and is $(2j_0+2j_0'+j_1+j_1')/x_0$
in the time direction.  The average electric field is easily worked out from the potential jump across the cell in the vertical direction, and is
$(V_1-V_5)/t_0$.
The ratio of the average current field divided by the average electric field is
\beq 
\begin{aligned}
\frac{(2j_0+2j_0'+j_1+j_1')/x_0}{(V_1-V_5)/t_0}=-\frac{t_0(j_0+j_0')[2+(\Gg_2/\Gg_1)]}{x_0(2\Gg_1j_0+2\Gg_2j_2)}
&=-\frac{t_0(j_0+j_0')[2+(\Gg_2/\Gg_1)]}{x_0(\Gg_1j_0+\Gg_2j_0+\Gg_1j_0'+\Gg_2j_0')}\\&=-\frac{t_0[2+(\Gg_2/\Gg_1)]}{x_0(\Gg_1+\Gg_2)}
\end{aligned}
\eeq{1.15a}
and this corresponds to the effective conductivity coefficient in the time direction, that is, $-\Gb_*$.

\subsection{Antisymmetric dynamics}

To solve for the currents and potentials associated with the antisymmetric dynamics with periodic currents (and hence periodic electric fields), we assign the convention
that currents flowing in the direction of positive time have positive sign, and currents flowing in the direction of negative time have negative sign. The currents
are labelled $j_0$, $j_1$, $j_2$ and $j_0'$, $j_1'$, $j_2'$ on the lines in \fig{6}.
\begin{figure}[!ht]
\centering
\includegraphics[width=0.7\textwidth]{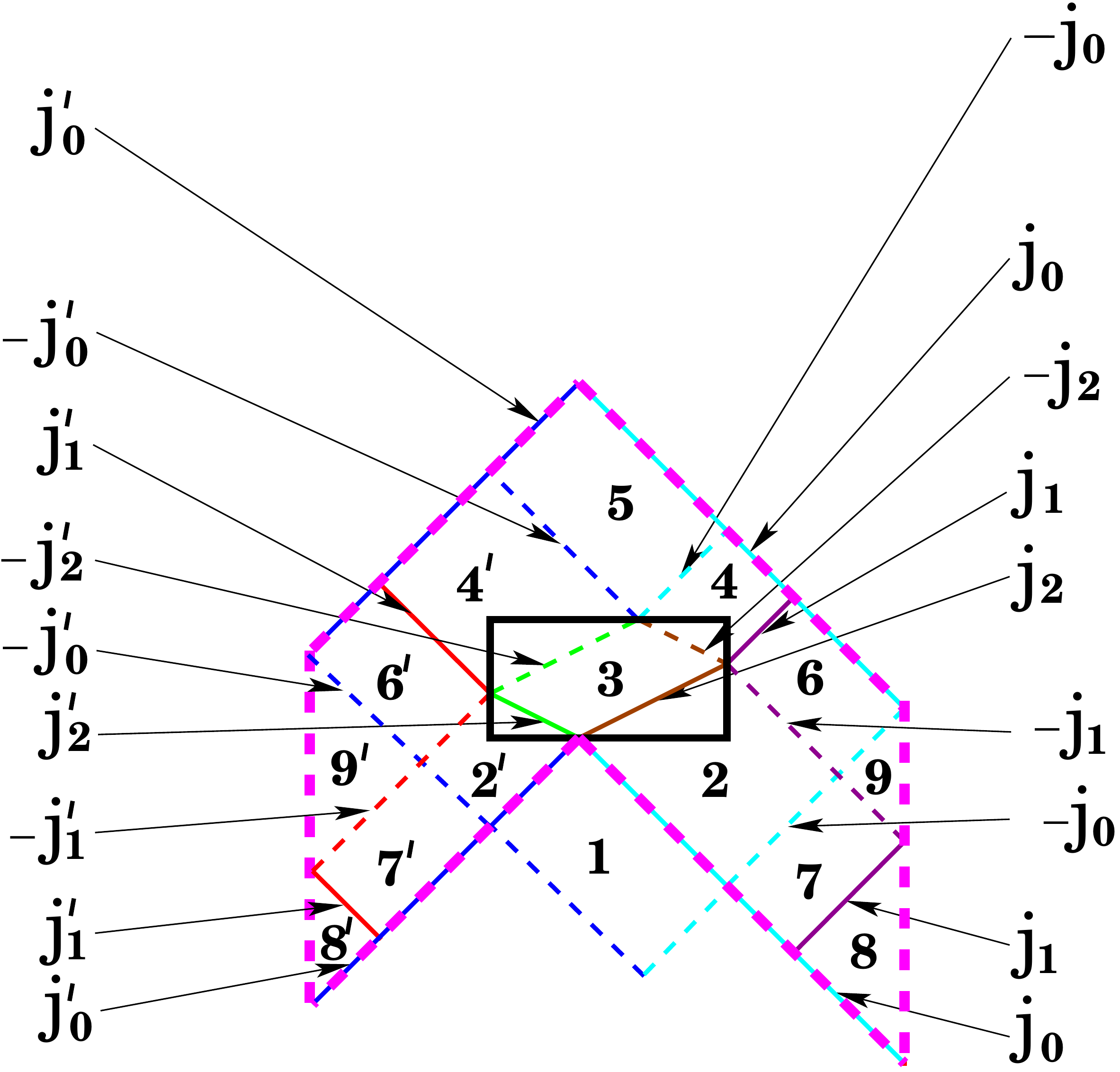}
\caption{The current flows in one half of the unit cell for antisymmetric dynamics. Here different colors are used to distinguish lines that have different currents flowing along them,
and dashed lines indicate currents that are measured with respect to decreasing time: thus a current $j_i$ flowing in the direction of decreasing time corresponds to a current $-j_i$
in the direction of increasing time. Periodicity and antisymmetry ensure that the currents flow as described in this figure.}
\labfig{6}
\end{figure}
Periodicity
and antisymmetry ensure the current flow has this structure. By conservation of current at the interfaces between the phases (see equation \eq{Continuity_flux}) we require that
\beq j_0+j_0'=j_2+j_2', \quad j_1=j_2, \quad j_1'=j_2'. \eeq{1.16} 
We let $V_i$, $i=1,2,\ldots, 9$ denote the potentials in each of the regions $1,2,\ldots, 9$ and we let $V_i'$, $i=2,4,6,7,8,9$ denote the potentials in each of the regions
$2',4',6',7',8',9'$. Once again, for simplicty, we set $V_1=0$. Then, from the rule that the potential jump across a characteristic line is $\Gg_i$ times the current flowing through it, we obtain
successively
\beq
\begin{aligned} 
&V_2 = \Gg_1j_0, \quad V'_2=\Gg_1j_0',\quad V_3=\Gg_1j_0+\Gg_2j_2=\Gg_1j_0'+\Gg_2j_2', \quad V_6=\Gg_1(j_0-j_1),\\& V_6'=\Gg_1(j_0'-j_1'), \quad
V_4 = \Gg_1j_0,\quad V_4'=\Gg_1j_0',\quad V_5=0,\quad V_7=0, \quad V_7'=0, \\& V_8  =  -\Gg_1j_1, \quad V_8'=-\Gg_1j_1', \quad V_9=-\Gg_1j_1,\quad  V_9'=-\Gg_1j_1'.
\end{aligned}
\eeq{1.17}
So in addition to \eq{1.12} the currents must satisfy the constraint
\beq \Gg_1j_0+\Gg_2j_2=\Gg_1j_0'+\Gg_2j_2'. \eeq{1.18}
Thus $j_0$, and $j_0'$ can be chosen independently, and in terms of these
\beqa  j_1=j_2& = &\frac{1}{2}(j_0+j_0')-\frac{\Gg_1}{2\Gg_2}(j_0-j_0'), \quad \nonum
      j_1'=j_2'& = &\frac{1}{2}(j_0+j_0')+\frac{\Gg_1}{2\Gg_2}(j_0-j_0').
\eeqa{1.19}
As $V_5=V_1$ there is no average electric field in the vertical direction. Similarly, there is no current field in the vertical direction. In the horizontal direction,
by looking at the current flux out of the unit cell, one sees the average current to the right is $2(j_1-j_0)/t_0$. The average electric field pointing to the right
is $(V_9'-V_9)/x_0$. The ratio of the average current field divided by the average electric field is then
\beq \frac{2(j_1-j_0)/t_0}{(V_9'-V_9)/x_0}=\frac{2x_0(j_1-j_0)}{t_0\Gg_1(j_1-j_1')}=\frac{x_0(j_0-j_0')(\Gg_1+\Gg_2)}{t_0\Gg_1^2(j_0-j_0')}=
\frac{x_0(\Gg_1+\Gg_2)}{t_0\Gg_1^2},
\eeq{1.20}
and this corresponds to the effective conductivity coefficient in the space direction, $\Ga_*$. 

Therefore, from \eq{1.8a}, \eq{1.15a}, and \eq{1.20} we see that the "effective conductivity tensor" is
\beq \BGs_*=\!\bpm \Ga_* & 0 \\ 0 & -\Gb_* \epm
\!=\! \bpm \frac{x_0(\Gg_1+\Gg_2)}{t_0\Gg_1^2}\! & 0\\ 0 &\!\! -\frac{t_0[2+(\Gg_2/\Gg_1)]}{x_0(\Gg_1+\Gg_2)} \epm
\!=\!\bpm \frac{c_1(c_1+2c_2)(\Gg_1+\Gg_2)}{\Gg_1^2(c_1+c_2)}\! & 0 \\ 0 & \!\!\!-\frac{(c_1+c_2)[2+(\Gg_2/\Gg_1)]}{c_1(c_1+2c_2)(\Gg_1+\Gg_2)}\epm.
\eeq{1.21}
The associated "effective speed" is therefore
\beq c_*=\sqrt{\Ga_*/\Gb_*}=\frac{c_1(c_1+2c_2)(\Gg_1+\Gg_2)}{c_1+c_2}\sqrt{\frac{1}{\Gg_1(2\Gg_1+\Gg_2)
}}.
\eeq{1.22}
Interestingly, this does not approach the wave velocity $c_1$, even when the parameters of phase $2$ approach those of phase $1$.
If we had chosen our units of space and time so that $x_0=1$ and $t_0=1$ then the corresponding dimensionless speed would be
\beq \frac{c_*t_0}{x_0}=(\Gg_1+\Gg_2)\sqrt{\frac{1}{\Gg_1(2\Gg_1+\Gg_2)
}}. \eeq{1.22a}
We have put quotes around "effective conductivity tensor" and "effective speed"  because at this stage it is unclear what is their physical significance. If the conductivity tensor field
$\BGs(\Bx)$ had a very tiny imaginary part (as relevant to composites of hyperbolic materials, when $\BGs(\Bx)$ is replaced by the dielectric tensor field $\BGve(\Bx)$, and time is replaced
by a spatial variable) then indeed $\BGs_*$ would be the appropriate "speed" giving the effective "characteristic lines" of propagation. In the absence of such an imaginary part one needs
to derive the appropriate homogenized equations. This homogenized equation is unlikely to be simply $\Div\BGs_*\Grad \underline{V}=0$ where $\underline{V}(\Bx)$ is some coarse grained potential,
as it is quite evident from the branching nature of field patterns (see \fig{1} and \fig{2}) that there should be some term giving dispersion in the effective equations.

\section{Associated field patterns}\label{Linear_potential}
\setcounter{equation}{0}
As before let us assume the origin $x=t=0$ coincides with the bottom left corner of an inclusion, as in \fig{1}. Now given a
non-negative index parameter $\phi<c_2$, suppose we launch a field pattern
$V(x,t)$ by setting the initial conditions
\beq V(x,0)=H(x-\phi),\quad j_2(x,0)=-\sqrt{\Ga_2\Gb_2}\Gd(x-\phi), \eeq{ass.1}
so that initially the discontinuity just propagates to the right: in the inclusion containing the origin, the potential $V(x,t)$ is just equal to $H(x-\phi-c_2t)$
until the time $1-\phi/c_2)$ when the discontinuity strikes the right hand side of the inclusion. Let us label this potential $V(x,t,\phi)$ 
and its associated current $\Bj(x,t,\phi)$ to make explicit
the dependence on $\phi$. Now given two index parameters $\phi_1$ and $\phi_2$, where $0<\phi_1<\phi_2<c_2$, we can consider the
potential 
\beq W(x,t,\phi_1,\phi_2)=\int_{\phi_1}^{\phi_2}V(x,t,\phi)~d\phi. \eeq{ass.2}
This function $W(x,t,\phi_1,\phi_2)$ is clearly a solution of the wave equation being a superposition of solutions. It has discontinuities in slope along
the lines indicated in \fig{ass1}.
\begin{figure}[!ht]
\centering
\includegraphics[width=0.7\textwidth]{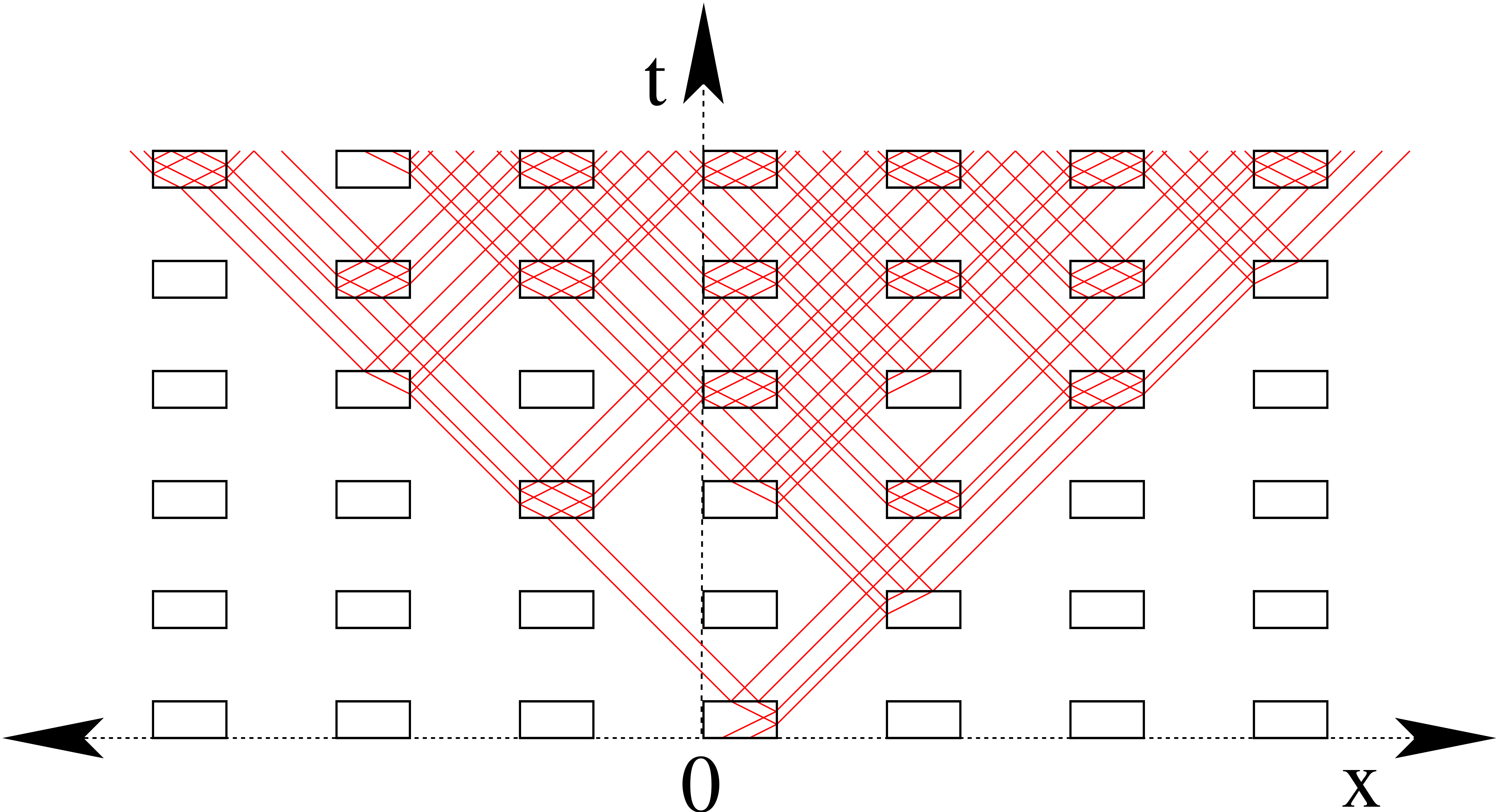}
\caption{Evolution of the lines of discontinuity of slope of the associated field pattern $W(x,t,\phi_1,\phi_2)$}
\labfig{ass1}
\end{figure}
Let the associated current field be labelled
$\BJ(x,t,\phi_1,\phi_2)$. It is given by 
\beq \BJ(x,t,\phi_1,\phi_2)=-\BGs(x,t)\Grad W(x,t,\phi_1,\phi_2)
=\int_{\phi_1}^{\phi_2}\Bj(x,t,\phi)~d\phi. \eeq{ass.3}
Clearly then, $W(x,t)$ satisfies the initial conditions,
\beqa W(x,0,\phi_1,\phi_2)=\int_{\phi_1}^{\phi_2}H(x-\phi)~d\phi
                                         & = &  0\quad {\rm if}~~x<\phi_1 \nonum
                                         & = & x-\phi_1 \quad {\rm if}~~\phi_1<x<\phi_2 \nonum
                                         &= & \phi_2-\phi_1\quad {\rm if}~~x>\phi_2, \nonum
J_2(x,0,\phi_1,\phi_2)=-\sqrt{\phi_2\Gb_2}\int_{\phi_1}^{\phi_2}\Gd(x-\phi)~d\phi 
                                         & = &  0\quad {\rm if}~~x<\phi_1 \nonum
                                         & = & -\sqrt{\phi_2\Gb_2} \quad {\rm if}~~\phi_1<x<\phi_2 \nonum
                                         &= & 0 \quad {\rm if}~~x>\phi_2,
\eeqa{ass.3a}
By considering the evolution of this disturbance, it is clear that the associated field pattern $W(x,t,\phi_1,\phi_2)$ is a linear function
of $x$ and $t$ in each space-time polygonal region with boundaries marked by the lines of discontinuity in slope in \fig{ass1}. Thus
the lines of discontinuity in the associated field pattern can be considered as  sources and sinks of the "electric field" 
$-\Md W(x,t,\phi_1,\phi_2)/\Md x$, like electrical charges. We call $W(x,t,\phi_1,\phi_2)$ an associated field pattern of the first degree.
An example of an associated field pattern of the first degree that is periodic is graphed in \fig{linear}
\begin{figure}[!ht]
\centering
\includegraphics[width=0.7\textwidth]{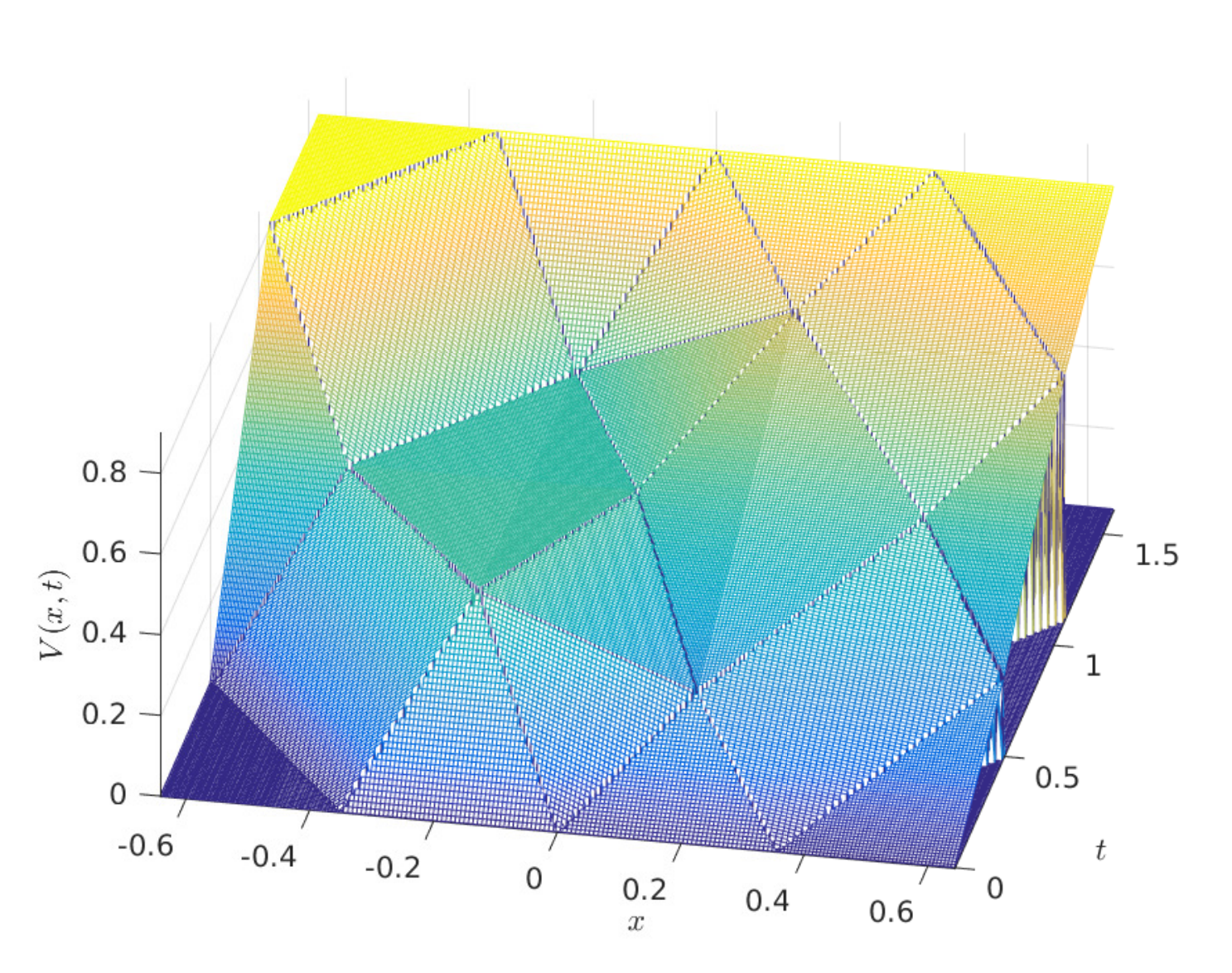}
\caption{An associated field pattern, with extreme values $\phi_1=0$ and $\phi_2=c_2$, that is generated from the symmetric periodic potentials $V(x,t,\phi)$ 
given in Section 3.2, with parameter choices $\alpha_1=0.5, \beta_1=1.5, \alpha_2=1$, and $\beta_2=8$, corresponding to $\gamma_1=1$ and $\gamma_2=3$,
and with currents $j_0=0.5$, and $j_0'=1$. Note the piecewise linear nature of the potential. The spurious vertical lines at the right back side are a numerical artifact and should be ignored.}
\labfig{linear}
\end{figure}

An associated field pattern of the second degree would be the function 
\beq Y(x,t,\phi_{11},\phi_{12},\phi_{21},\phi_{22})=\int_{\phi_{11}}^{\phi_{12}}d\phi_1\int_{\phi_{21}}^{\phi_{22}}d\phi_2~W(x,t,\phi_1,\phi_2).
\eeq{ass.4}
where $\phi_{11},\phi_{12},\phi_{21},$ and $\phi_{22}$ are parameters such that 
\beq 0<\phi_{11}<\phi_{12}<\phi_{21}<\phi_{22}<c_2. \eeq{ass.5}
This function $Y(x,t,\phi_{11},\phi_{12},\phi_{21},\phi_{22})$ will have discontinuities in its second derivatives across the lines in the field pattern.
Field patterns of higher degree can then be defined in the obvious way.

\section{Numerical results}\label{Numerical_result_alig}
\setcounter{equation}{0}

To test the theoretical results obtained in Section \ref{Symm_section} and explore the subject of field patterns
further, we perform some numerical analyses. Since a field pattern lives on its own discrete network, it is only
necessary to study the dynamics of field patterns on these discrete networks. On each discrete network the 
evolution of the currents in the ``wires" can be viewed as a dynamical system, and due to the periodicity in time
it suffices to study the currents in the ``wires'' at times $t=\tau+n\,t_0$ for $n=0,1,2,...$, where $\tau$ is some fixed time.
To simplify the description we choose $\tau$ to be a time where none of the``wires" in the discrete network
intersect, as shown for example by the horizontal line at $t=\tau$ in \fig{3}.
Then the state of the system at these discrete times is entirely determined by specifying the currents in each of
the "wires": it is captured by the function $j(k,m,n)$ where $k$, taking integer values between $1$ and $12$ indexes the "wires" in each unit cell, the integer $m$ indexes the cell, and the integer $n$ indexes the discrete time.
The state at any intervening time between the discrete times is easily determined from the dynamics:
we can think of the set of currents at the discrete times as being representative of the full dynamics in much the
same way that a Poincar{\'e} map helps one visualize the dynamics of a dynamical system.

Our aim is to determine the evolution of the state function $j(k,m,n)$, as $n=0,1,2,...$, which is representative of the time, increases. To achieve such a goal, we calculate the Green function that allows one to recover the currents at a certain time $t=\tau+n\,t_0$ with $n$ fixed, given the currents at time $t=\tau+(n-1)\,t_0$. Clearly, this is done by 
taking one unit cell, say with $m=m_0$ and injecting, at $t=\tau$ ($n=0$), a unitary current in each of the 12 points $k=1,2,\ldots,12$ marked by the green dots on the line $t=\tau$ in \fig{3}, one at a time, and by calculating how such a current flows along the characteristic lines to determine the currents at $t=\tau+t_0$ ($n=1$). Note that the current injected in some of the 12 points may cross the boundary of the unit cell: if one injects, for instance, a current at point 1, this will flow towards the points 9, 10 and 12 of the adjacent cell on the left, having $m=m_0-1$, whereas, if one injects current at point 12, for example, this will flow towards the points 1, 3 and 4 of the adjacent cell on the right, having $m=m_0+1$. This means that the currents injected in a unit cell at $t=\tau$ ($n=0$) may influence at $t=\tau+t_0$ ($n=1$), the currents in up to three unit cells. 

The Green function so calculated, then, can be denoted by $G_{k,k'}(m-m')$ to indicate that it provides the current at point $k$, with $k=1,2,...,12$, of cell $m$, given the current at point $k'$, with $k'=1,2,...,12$, of cell $m'$. Such a function obviously only depends on the differences $m-m'$, and its explicit expression is given in Section \ref{Green_function}. Then, the current at the point $k$ of cell $m$ at time $t=\tau+n\,t_0$ is determined by the currents at points $k'$ of cells $m'$ at the previous time $t=\tau+(n-1)\,t_0$ by 
\beq j(k,m,n)=\sum_{k',m'}T_{(k,m),(k,'m')}j(k',m',n-1), \eeq{3.1}
where
\beq T_{(k,m),(k,'m')}=G_{k,k'}(m-m') \eeq{3.2}
is the transfer matrix.
To approximate the hypothesis of an infinite medium in the $x$ direction we could consider a very large number $M$ of cells. It is then natural
to take periodic boundary conditions, so that we can think of the discrete network as lying on the surface of a cylinder, with the axis variable
corresponding to time and the angle variable corresponding to space. Thus in \eq{3.1} cell $M$ is identified with cell $0$ and cell $-1$ is identified with
cell $M-1$: equivalently the argument $m-m'$ of $G_{k,k'}(m-m')$ in \eq{3.1} should be replaced with $(m-m')\mod M$.

\subsection{On testing the periodic dynamics}

To test the periodic solution corresponding to the symmetric dynamics that we derived in the previous section, we have just to impose that, at $t=\tau$, the distribution of currents in the 12 points of each unit cell be equal to the periodic distribution of currents given by the symmetric dynamics of \fig{4}, and check that for $n=1,2,...$, such a distribution does not change. In \fig{num_periodic} we represent the time evolution of the currents for $n=0,1,..,100$ in 10 cells (so that the total number of points is 120), for the case when $j_0=1$ and $j_0'=-2$, and $\Gg_1=1$ and $\Gg_2=3$. As expected, the solution depicted in \fig{num_periodic} is clearly periodic both in space and time.  

\begin{figure}[!ht]
\centering
\includegraphics[width=0.7\textwidth]{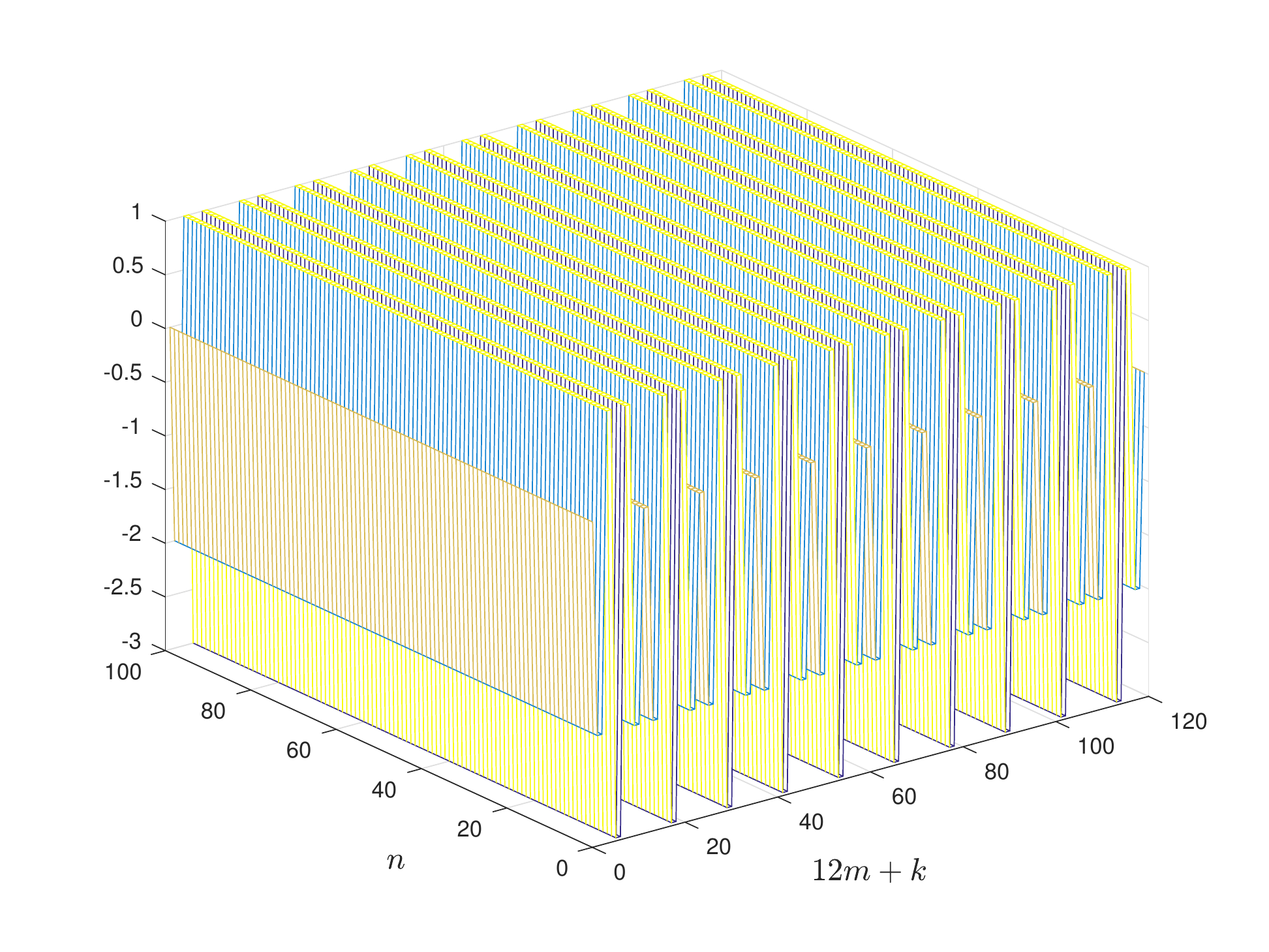}
\caption{Evolution of the currents flow in 10 cells (since there are 12 test points in each cell, the total number of points monitored is equal to 120), for $t=\tau+nt_0$ with $n=0,1,..,100$, in the case when the distribution of currents injected at $t=\tau$ ($n=0$) is equal to the periodic distribution given by the symmetric dynamics of \fig{4}, with $j_0=1$ and $j_0'=-2$, and $\Gg_1=1$ and $\Gg_2=3$.}
\labfig{num_periodic}
\end{figure}
A similar result is obtained when one considers as initial distribution of currents that corresponding to the antisymmetric dynamics of \fig{6}.
 

\subsection{Blow-up}

At this point one may ask what happens when at $t=\tau$ the current is injected only at one of the 120 points. In order to address this question, we supposed that at time $t=\tau$ the only non-zero current is the one at point 1 of cell 5, that is, at the point labeled with $61$. It is obvious that, due to the periodicity condition for which cell 1 is treated as adjacent to cell 10, the point of injection of the current could be the point 1 of any cell. We found that the solution blows up exponentially with time.
In order to better appreciate such a behavior, in \fig{num_logc100} we report the solution for the case of 100 cells with the current injected only at point $601$, that is, point 1 of cell 50. 

\begin{figure}[!ht]
\centering
\includegraphics[width=0.7\textwidth]{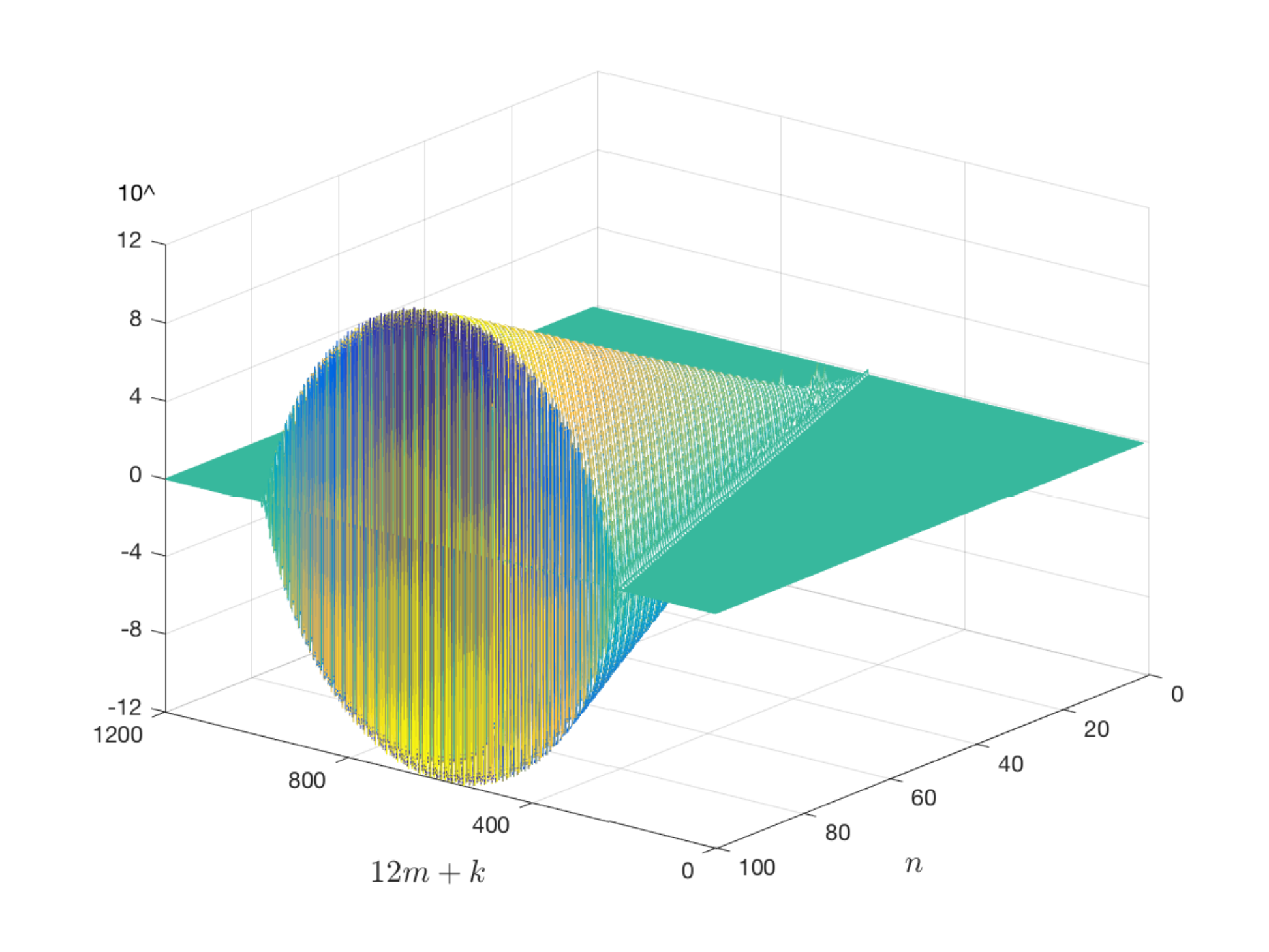}
\caption{Evolution of the currents flow in 100 cells (1200 points in total), for $t=\tau+nt_0$ with $n=0,1,..,100$, in the case when a unitary current is injected only at point 1 of cell 50, that is, at the point labeled with 601. The solution, here represented in logarithmic scale, blows up exponentially.}
\labfig{num_logc100}
\end{figure}

One way to avoid this blow-up is to set $\Gg_1=\Gg_2$, so there is no impedance mismatch, as done in the analysis of  Lurie \cite{Lurie:1997:EPS}.
Then, if we start by injecting current at a single point, the field pattern degenerates to a single trajectory.  The current flows either to the right or to the left in the space-time diagram: there are no reflected waves but only a transmitted wave. Conservation of current then implies that the
current along the trajectory remains constant: there is no blow-up.

\subsection{Eigenvalues and Eigenvectors of the Transfer Matrix}

Now the question is: is it possible to obtain a solution that does not blow up without imposing any special constraint on the material parameters? To this purpose, let us calculate the eigenvalues of the transfer matrix. Clearly, with periodic boundary conditions, the transfer matrix is a $12M\times 12M$ matrix, and 
\eq{3.2} gives its elements in terms of the Green function, whose components are explicitly given in Section \ref{Green_function}.
 For simplicity, suppose we consider only 10 cells, for a total of 120 points (12 points for each unit cell) so that the number of eigenvalues is equal to 120 (we write the transfer matrix as a $120\times 120$ matrix that, applied to a vector with 120 components describing the current distribution in the 120 points at a certain time, provides the vector with 120 components representing the currents in the same 120 points after a period of time). For the particular case when $\Gg_1=1$ and $\Gg_2=3$, these eigenvalues are plotted in \fig{eigen}. From the graph it is clear that many eigenvalues are on the unit circle.
This is a consequence of the ${\cal P}{\cal T}$ symmetry of the system. Here ${\cal P}$ stands for parity symmetry, the reflection invariance of the microstructure under spatial
reflection $x\to -x$ when the origin is chosen at the center of an inclusion, and ${\cal T}$ stands for time symmetry, the reflection invariance of the microstructure under time
reversal $t\to -t$ (when again the origin is chosen at the center of an inclusion). Excellent reviews of  ${\cal P}{\cal T}$ symmetry and its application to optics are given by
Suchkov, Sukhorukov, Huang, Dmitriev, Lee, and Kivshar \cite{Suchkov:2016:NSS} and Konotop, Yang and Zezyulin \cite{Konotop:2016:NWP}. It may seem that 
the time reversal symmetry is broken in our discrete dynamics, by the choice of $\tau$. Note, however, that nothing really changes if we modify $\tau$ so long as the line
$t=\tau$ does not intersect the inclusions: one just has to label the currents appropriately so they are consistent at the different values of $\tau$. Thus if the line $t=\tau$ moves
through a point where two current lines cross, then the ordering of the current numbering labels undergoes a swap. One can even choose $\tau$ so the line $t=\tau$ is exactly midway
between two rows of inclusions, and then one clearly has time reversal symmetry. Nothing really changes, but at the points of intersection of two current lines one has to be careful to
distinguish the current flowing upwards to the right from the current flowing upwards to the left. 
\begin{figure}[!ht]
\centering
\includegraphics[width=0.7\textwidth]{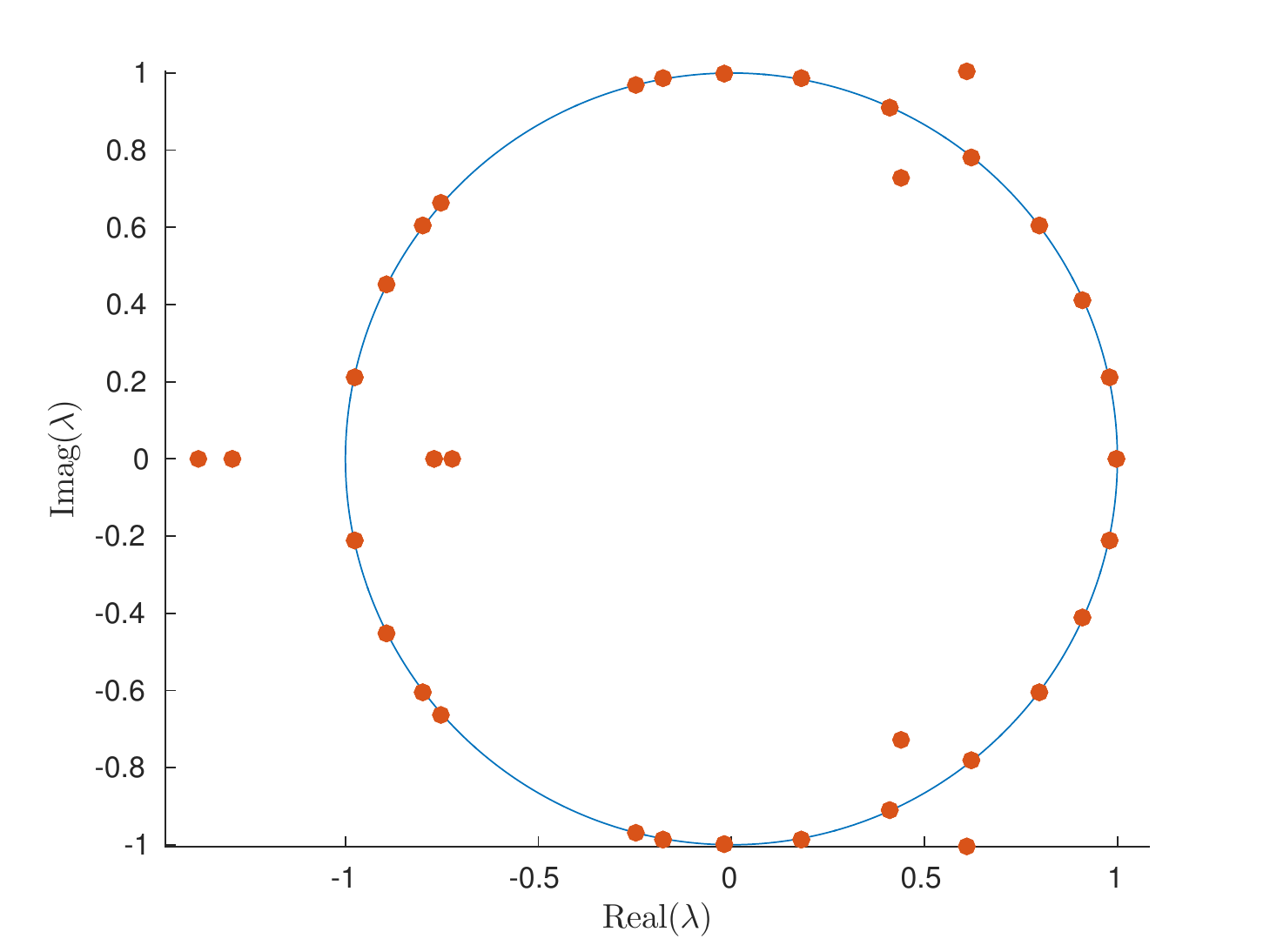}
\caption{Eigenvalues of the transfer matrix in the case $\Gg_1=1$ and $\Gg_2=3$.}
\labfig{eigen}
\end{figure}

Clearly modes having eigenvalues with modulus greater than 1 blow up in time, those with eigenvalues having modulus  less than 1 decay in time, while those with eigenvalue having modulus 1 oscillate with time.
Among all the eigenvalues, we select those with modulus equal to 1, and we apply the corresponding eigenvector as the initial current distribution in each unit cell. If $\Gg_1=1$ and $\Gg_2=3$ there are 99 eigenvalues with modulus 1 in total: the real one has multiplicity 3, whereas among the 96 complex eigenvalues the independent ones sum up to 13 pairs (each pair includes one eigenvalue together with its complex conjugate). In particular, 7 couples have negative real part and 6 positive real part. The results can be grouped into 3 classes: solutions periodic both in time and space with time period equal to $t_0$ and space period equal to $2x_0$, solutions periodic both in time and space but with period larger than $t_0$ and $2x_0$, and finally, solutions that are not periodic. To the first class belong the symmetric (\fig{4}) and antisymmetric (\fig{6}) dynamics considered in the two previous subsection and, therefore, we refer to \fig{num_periodic}. Solutions of the second type are shown, as an example, in \fig{num_eig51_52}, whereas solutions of the third type are given, for instance, in Figures \figs{num_eig84_85} and \figs{num_eig86_87}.

\begin{figure}[!ht]
\centering
\includegraphics[width=0.7\textwidth]{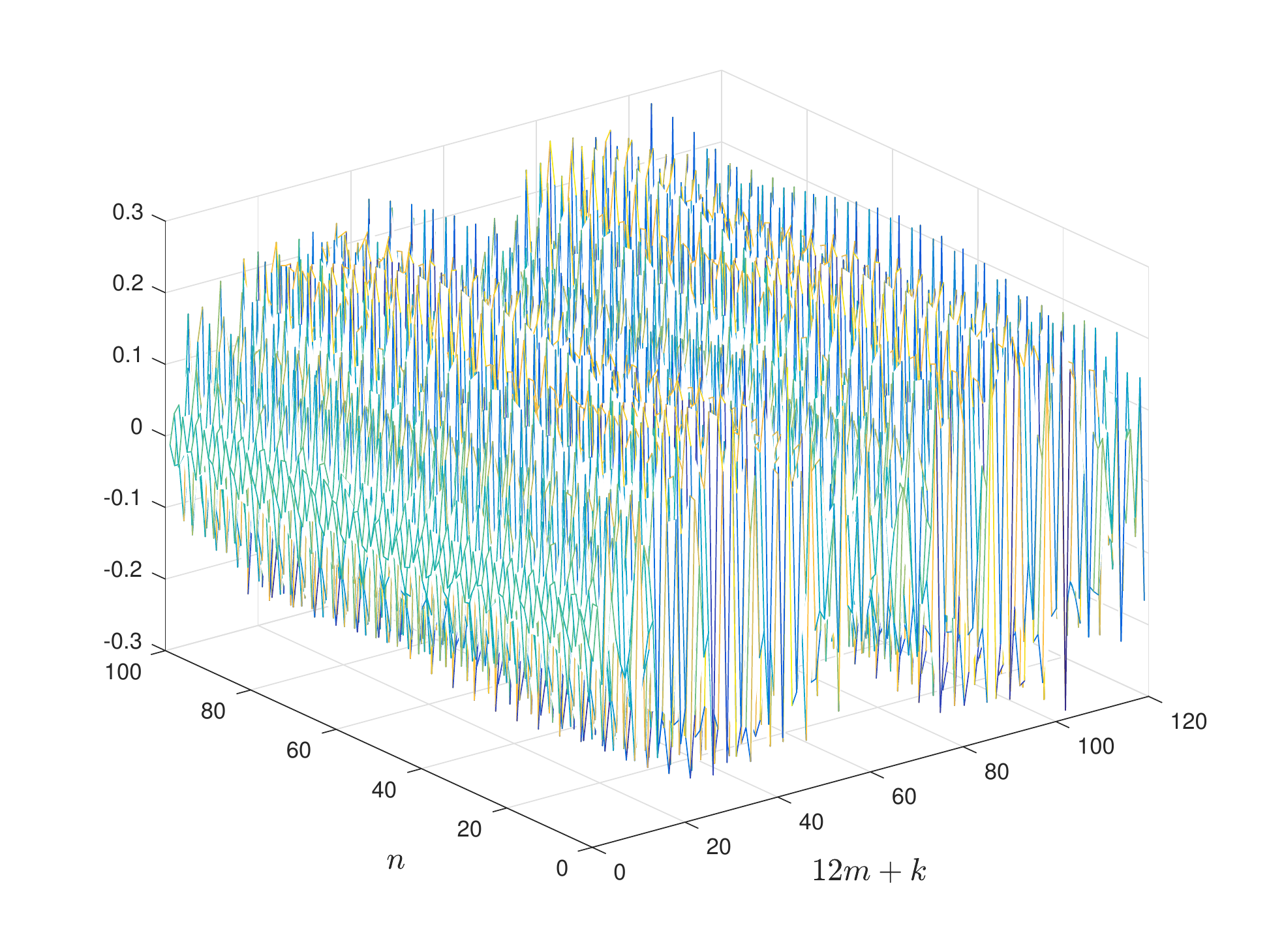}
\caption{Evolution of the currents flow in 100 cells for $t=\tau+nt_0$ with $n=0,1,..,100$, in the case when $\Gg_1=1$ and $\Gg_2=3$ and, at $t=\tau$ (n=0), we inject a distribution of currents equal to that given by the sum of one of the couples of two conjugate eigenvectors corresponding to the eigenvalues $-0.9775\pm 0.211$i (with multiplicity 4). The current flows following a pattern that is periodic both in time and in space but the periodicity is not equal to that of the unit cell.}
\labfig{num_eig51_52}
\end{figure}
\begin{figure}[!ht]
\centering
\includegraphics[width=0.7\textwidth]{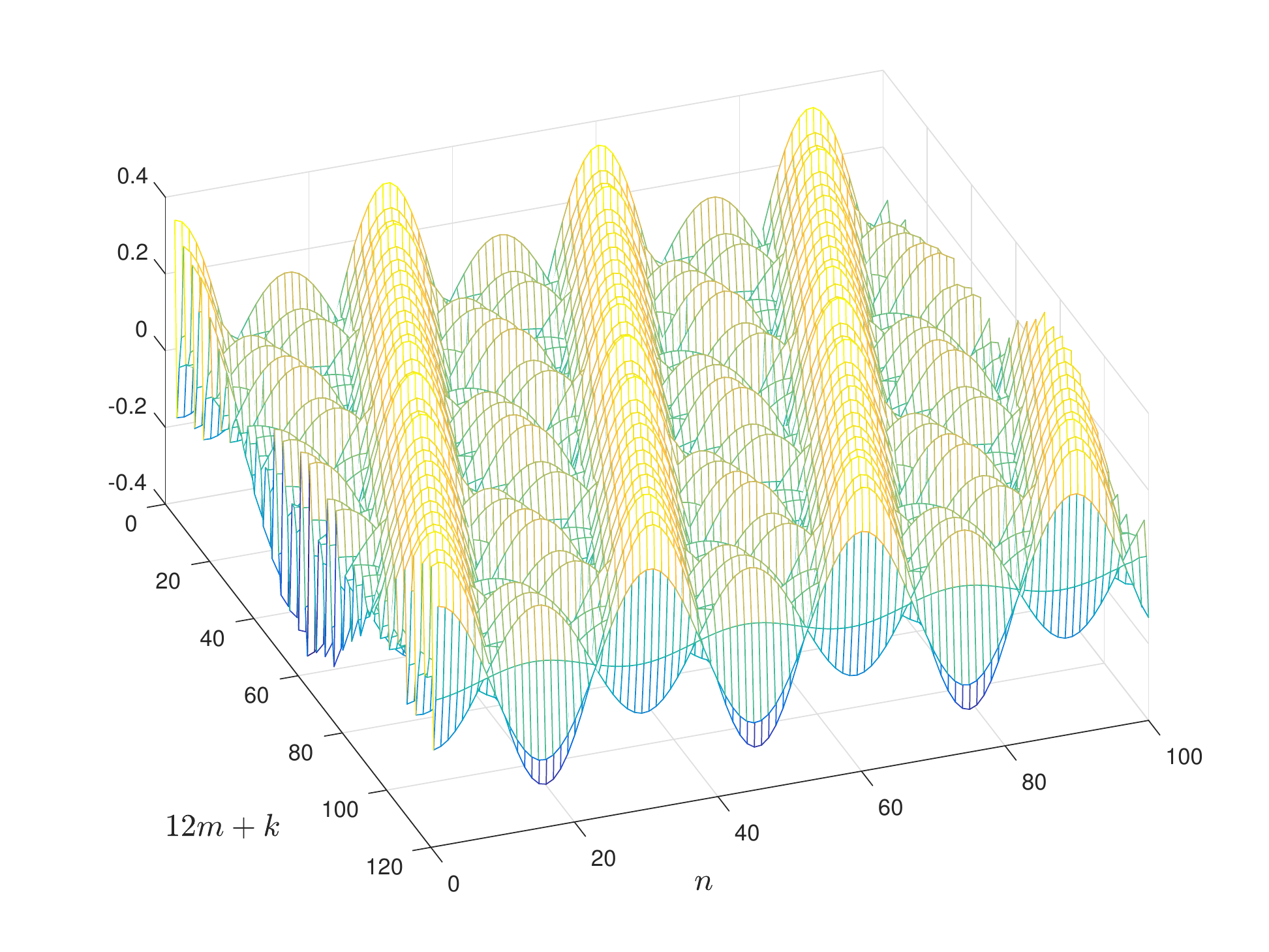}
\caption{Evolution of the currents flow in 100 cells for $t=\tau+nt_0$ with $n=0,1,..,100$, in the case when $\Gg_1=1$ and $\Gg_2=3$ and, at $t=\tau$ (n=0), we inject a distribution of currents equal to that given by the sum of one of the couples of two conjugate eigenvectors corresponding to the eigenvalues $0.9775\pm 0.211$i (with multiplicity 4). The current flows following a pattern that is not periodic.}
\labfig{num_eig84_85}
\end{figure}
\begin{figure}[!ht]
\centering
\includegraphics[width=0.7\textwidth]{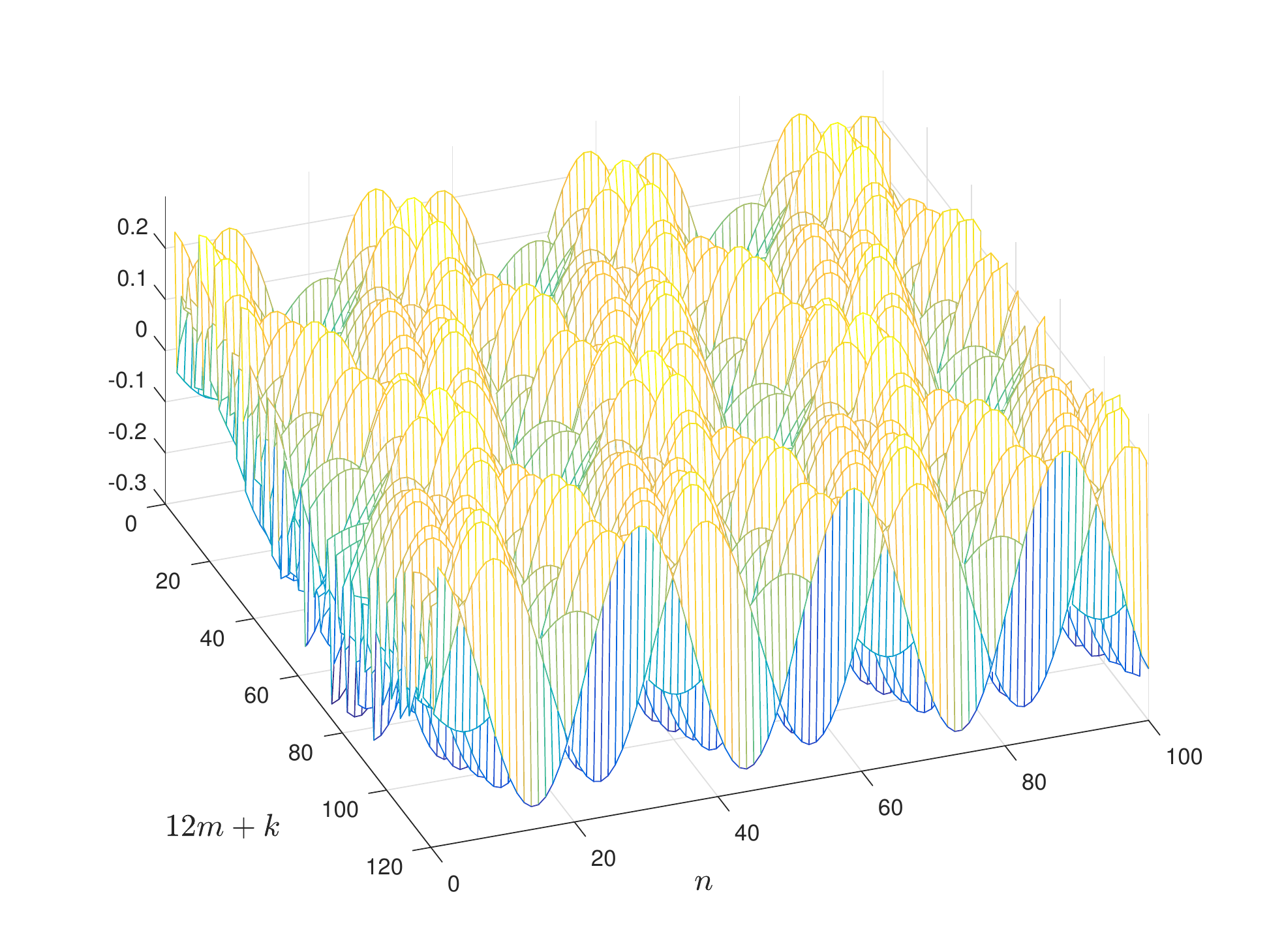}
\caption{Evolution of the currents flow in 100 cells for $t=\tau+nt_0$ with $n=0,1,..,100$, in the case when $\Gg_1=1$ and $\Gg_2=3$ and, at $t=\tau$ (n=0), we inject a distribution of currents equal to that given by the sum of one of the couples of two conjugate eigenvectors corresponding to the eigenvalues $0.9775\pm 0.211$i (with multiplicity 4). The current flows following a pattern that is not periodic.}
\labfig{num_eig86_87}
\end{figure}

\section{Conclusions}
\setcounter{equation}{0}


This paper launches the study of field patterns, but leaves many avenues for further research. In particular, it should motivate subsequent investigations on how things change if there is a small
non-linearity (of possible relevance to quantum mechanics); how things change if the tensor $\BGs(\Bx)$ has a small imaginary part (of possible relevance to determining the effective behavior of 
composites of hyperbolic metamaterials, when $\BGs(\Bx)$ is replaced by the dielectric tensor field $\BGve(\Bx)$  and time is replaced by a spatial variable); how things change if one considers the wave equation in two or three spatial dimensions, rather than just one; how things change if one looks at field patterns associated with other equations such as Maxwell's equations with a space time microstructure, or perhaps a modified version of Dirac's equation that allows one to insert some space time microstructure.  A key point is that the underlying wave equation, in the ideal case, should not have any dispersion since
otherwise the field patterns will lose their structure.
Also our analysis begs the question as to whether there are space-time microstructures that give rise to field patterns, with appropriate moduli such that the
transfer matrix only has eigenvalues with modulus 1, so that there are no growing modes. Additionally, from the viewpoint of homogenization, and high frequency homogenization 
\cite{Bensoussan:1978:AAP,Birman:2006:HMP,Brassart:2010:ABE,Craster:2010:HFH,Allaire:2011:DGO,Hoefer:2011:DMH,Ceresoli:2015:DEA,Harutyunyan:2016:HFH}
in particular, one would like to know if there are solutions that have rapid oscillations at the scale of the cells, but which have macroscopic modulations and one would want to describe the effective equations that describe how these
modulations propagate. It will be exciting to see how our understanding develops. 

\section*{Acknowledgements}
The authors are grateful to the Minneapolis Institute for Mathematics and its Applications for support as part
of the special year on Mathematics and Optics. They also thank the National Science Foundation for support through grant DMS-1211359. Alexander and Natasha Movchan, and Hoai-Minh Nguyen are thanked for comments on the manuscript. Maxence Cassier is thanked for suggesting that we plot the eigenvalues of the transfer matrix. Carme Calderer is thanked for interesting discussions about liquid crystals with space-time microstructures.

\section{Appendix A: Green's function for the space-time microstructure with aligned geometry}\label{Green_function}
\setcounter{equation}{0}

In this section we give the components of the Green function associated with the space-time microstructure illustrated in \fig{1}, where the rectangular inclusions are aligned. We inject unitary current at time $t=\tau$ at
each of the 12 points marked by the 12 green dots on the line $t=\tau$ in \fig{3} and
recall that at each time step currents injected into one cell, can generate currents at time $t=\tau+t_0$
not only in that cell, but also in the two neighboring cells. The Green function is calculated by determining how the currents split along the characteristics. Clearly, as the Green function only depends on $m-m'$, the case where the currents are injected at points in other cells is straightforward: one has just to suitably translate the expressions of the components of the Green function. Recall that  $G_{k,k'}(m-m')$ gives the current at point $k$, with $k=1,2,...,12$, of cell $m$, given the current at point $k'$, with $k'=1,2,...,12$, of cell $m'$. Since this only depends on $m-m'$ [more precisely $(m-m')$ mod $M$] it suffices to take $m'=0$.  
Then $G_{k,k'}(m)$ gives the current at point $k$, with $k=1,2,...,12$, of cell $m$, given the current at point $k'$, with $k'=1,2,...,12$, of cell $0$: $m=-1$ then refers to the cell on the left of cell $0$, i.e. cell $M-1$, and $m=+1$ refers
to the cell on the right of cell $0$, i.e. cell $1$. So we see, step by step, if the current is injected at
\beq\nonumber
\begin{aligned}
&\text{Point 1 of cell 0}: j(1,2,0)=1\quad\Rightarrow\quad \left\{\begin{aligned} &G(9,1,-1)=1; \, G(10,1,-1)=\frac{\Gg_1-\Gg_2}{\Gg_1+\Gg_2};\\&\,G(12,1,-1)=-\frac{\Gg_1-\Gg_2}{\Gg_1+\Gg_2}\end{aligned}\right.\\&
\text{Point 2 of cell 0}: j(2,2,0)=1\quad\Rightarrow\quad G(1,2,0)=-\frac{\Gg_1-\Gg_2}{\Gg_1+\Gg_2}; \, G(3,2,0)=\frac{\Gg_1-\Gg_2}{\Gg_1+\Gg_2};\,G(6,2,0)=1\\&
\text{Point 3 of cell 0}: j(3,2,0)=1\quad\Rightarrow\quad G(11,3,-1)=1\\&
\text{Point 4 of cell 0}: j(4,2,0)=1\quad\Rightarrow\quad G(8,4,0)=1\\& 
\text{Point 5 of cell 0}: j(5,2,0)=1\quad\Rightarrow\quad G(1,5,0)=1; \, G(4,5,0)=\frac{\Gg_1-\Gg_2}{\Gg_1+\Gg_2};\,G(6,5,0)=-\frac{\Gg_1-\Gg_2}{\Gg_1+\Gg_2}\\&
\text{Point 6 of cell 0}: j(6,2,0)=1\quad\Rightarrow\quad G(7,6,0)=-\frac{\Gg_1-\Gg_2}{\Gg_1+\Gg_2}; \, G(9,6,0)=\frac{\Gg_1-\Gg_2}{\Gg_1+\Gg_2};\,G(10,6,0)=1\\&
\text{Point 7 of cell 0}: j(7,2,0)=1\quad\Rightarrow\quad G(3,7,0)=1; \, G(4,7,0)=\frac{\Gg_1-\Gg_2}{\Gg_1+\Gg_2};\,G(6,7,0)=-\frac{\Gg_1-\Gg_2}{\Gg_1+\Gg_2}\\&
\text{Point 8 of cell 0}: j(8,2,0)=1\quad\Rightarrow\quad G(7,8,0)=-\frac{\Gg_1-\Gg_2}{\Gg_1+\Gg_2}; \, G(9,8,0)=\frac{\Gg_1-\Gg_2}{\Gg_1+\Gg_2};\,G(12,8,0)=1\\&
\text{Point 9 of cell 0}: j(9,2,0)=1\quad\Rightarrow\quad G(5,9,0)=1\\&
\text{Point 10 of cell 0}: j(10,2,0)=1\quad\Rightarrow\quad G(2,10,1)=1\\&
\text{Point 11 of cell 0}: j(11,2,0)=1\quad\Rightarrow\quad\left\{\begin{aligned}& G(7,11,0)=1; \, G(10,11,0)=\frac{\Gg_1-\Gg_2}{\Gg_1+\Gg_2};\\&\,G(12,11,0)=-\frac{\Gg_1-\Gg_2}{\Gg_1+\Gg_2}\end{aligned}\right.\\&
\text{Point 12 of cell 0}: j(12,2,0)=1\quad\Rightarrow\quad \left\{\begin{aligned}&G(1,12,1)=-\frac{\Gg_1-\Gg_2}{\Gg_1+\Gg_2}; \, G(3,12,1)=\frac{\Gg_1-\Gg_2}{\Gg_1+\Gg_2};\\&\,G(4,12,1)=1\end{aligned}\right.\\
\end{aligned}
\eeq{Green}
Obviously, all the other components are equal to zero.

\section{Appendix B: Solving the cell problem for a space-time microgeometry with staggered inclusions}
\setcounter{equation}{0}

In this Section we aim at deriving moduli that could partlially govern the effective behavior of the dynamic composite material having the 
staggered inclusion microstructure shown in 
Figure 4 of the main text. We use the same procedure used for the space-time geometry with aligned inclusions, illustrated in Figure 3 of the main text and presented in Section 3
of the main text. In particular, we consider the same initial conditions applied in Section 3 of the main text so that the potentials $V^+_i(x,t)$ and $V^-_i(x,t)$ and the corresponding currents are still described by equations (3.2) and (3.3), respectively. 

Note that the only difference between the microstructure with aligned inclusions (Figure 3 in the main text) and that with staggered inclusions (Figure 4 in the main text) consists in the fact that for the latter the unit cell of the dynamics network has the same dimensions of the unit cell of the geometry, that is $x_0$ and $t_0$. This also implies that the unit cell has no symmetries and, therefore, we cannot separate the behavior into symmetric and antisymmetric dynamics. However, we can still look for distributions of currents such that the only non-zero component of the average current (and average electric field) is either in the time direction or in the space direction as shown here in \fig{7} and \fig{8}, respectively. The unit cell can be taken to be that outlined by the dashed magenta lines in \fig{green}.

\begin{figure}[!ht]
	\centering
	\includegraphics[width=0.7\textwidth]{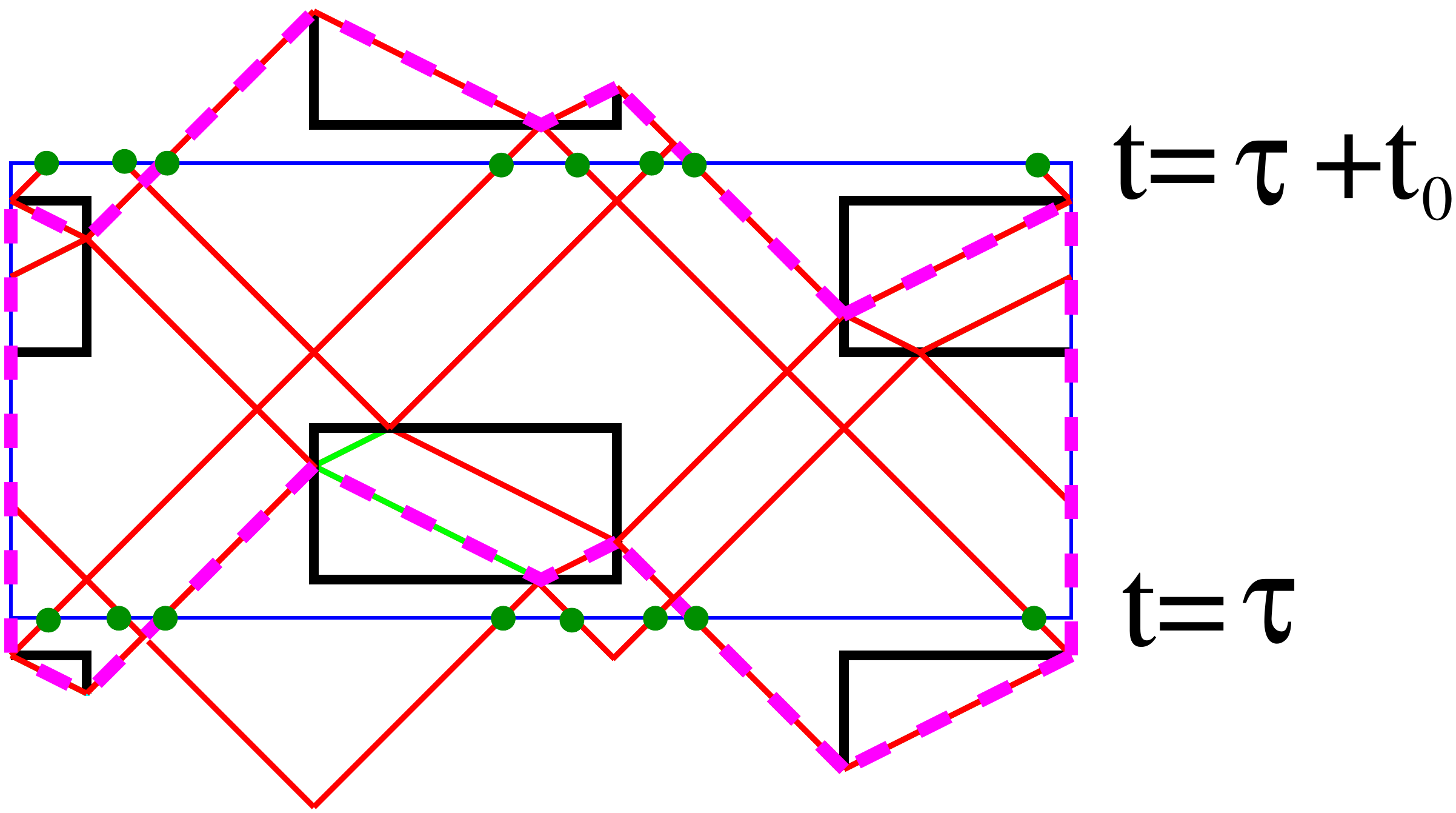}
	\caption{Outlined by the purple dashed lines is the unit cell of periodicity for the staggered cell dynamics of the microgeometry shown in Figure 4 of the main text in the case of a periodic field pattern.
		The red lines denote characteristic lines. The current flows along such lines and, across them, there is a jump in potential. Note that the dynamics 
		splits into dynamics that are symmetric and antisymmetric with respect to reflection about the vertical centerline of the cell. We call these the symmetric dynamics and the 
		antisymmetric dynamics. If the field pattern is not periodic then we can still consider a periodic array of the characteristic lines, and the period cell could be taken to be that
		outlined in blue. Current injected at time $t=\tau$ at any of the 8 points marked by the green dots at the bottom of the unit cell can contribute towards the excitation of the same field pattern. This current
		exits the unit cell at time $t=\tau+t_0$ at any of the 8 points  marked by the green dots at the top of the unit cell, and may flow to the exit points of adjoining unit cells as well.}
	\labfig{green}
\end{figure}

Let us start by considering the dynamics depicted here in \fig{7}: all the currents
are flowing in the direction of positive time, and have the magnitudes $j_0$, $j_1$, $j_2$, $j_3$, $j_4$, $j_5$, and $j_6$ as indicated on the lines in \fig{7}. Periodicity ensures the current flow has this structure. By conservation of current at the interfaces between the phases, we require that
\beq j_1+j_2=j_5+j_6. \eeq{Stagg_1} 

With reference to \fig{7} here, we let $V_i$, $i=1,2,\ldots, 12$ denote the potentials in each of the regions $1,2,\ldots, 12$ and we let $V_i'$, $i=1,3,5,6,\ldots,12$ denote the potentials in each of the regions
$1',3',5',6',\ldots,12'$. For simplicity, we set $V_1'=0$. Then, following the rule according to which the potential jump across a characteristic line is $\Gg_i$ times the current flowing through it, we obtain
\beq\begin{aligned} &V_2 =\Gg_2j_5,\quad V_3=\Gg_2(j_5+j_6), \quad V_3'=2\Gg_2j_5,\quad V_4=V_6=V_6'=\Gg_2(j_5+j_6)+\Gg_1j_1=2\Gg_2j_5+\Gg_1j_2,
	\\& V_5=2\Gg_2j_5+\Gg_1(j_1+j_2),\quad V_5'=2\Gg_2j_5+2\Gg_1j_2,\quad
	V_7= 2\Gg_2j_5+\Gg_1(j_2-j_3),\\&
	V_8=\Gg_2(2j_5+j_6)+\Gg_1(j_2-j_3), \quad V_8' =3\Gg_2j_5+\Gg_1(j_2-j_4),\quad V_9= \Gg_2(j_5+j_6)+\Gg_1(j_2-j_3),\\&
	V_{10} = V_{11}=\Gg_2(j_5+j_6)-\Gg_1j_3, \quad V_{10}'=V_{11}'=2\Gg_2j_5-\Gg_1j_4=\Gg_1j_4,\quad V_7'=2\Gg_2j_5+\Gg_1(j_2-j_4), \\&  
	V_{12} = \Gg_2(j_5+j_6)-\Gg_1(j_1+j_3),\quad V_{12}'=\Gg_1(j_4-j_2).
\end{aligned}
\eeq{Stag_2} 
Thus, in addition to \eq{Stagg_1}, we have two further constraints:
\beq \Gg_2(j_6-j_5)=\Gg_1(j_2-j_1), \quad \Gg_2j_5=\Gg_1j_4. \eeq{Stagg_3}
A fourth relation can be obtained by imposing $V_1=0$ and by evaluating the potential $V_{10}$ clockwise (as $V_{10}=2\Gg_2j_6-\Gg_1j_3$) and then anticlockwise ($V_{10}=\Gg_2j_3$), to give
\beq \Gg_2j_6=\Gg_1j_3 \eeq{Stagg_4}
So all the currents can be expressed in terms of only two independent currents, say $j_1$ and $j_2$. Then,
\beqa  j_5& = &\frac{1}{2}(j_1+j_2)-\frac{\Gg_1}{2\Gg_2}(j_2-j_1), \quad j_4=\Gg_2j_5/\Gg_1, \nonum
j_6& = &\frac{1}{2}(j_1+j_2)+\frac{\Gg_1}{2\Gg_2}(j_2-j_1), \quad j_3=\Gg_2j_6/\Gg_1.
\eeqa{Stagg_5}
As already pointed out, the current distribution shown in \fig{7} is such that the average current field is only in the time direction and it is easily worked out from the flux of current into the lower boundary of the cell of \fig{7}: $(2j_1+2j_2+j_3+j_4+2j_5+2j_6)/x_0$.  The average electric field can be easily determined from the potential jump across the cell in the vertical direction, and is $(V_1'-V_5')/t_0$.
The ratio of the average current field divided by the average electric field is then given by
\beq \frac{(2j_1+2j_2+j_3+j_4+2j_5+2j_6)/x_0}{(V_1'-V_5')/t_0}=-\frac{t_0(2j_1+2j_2+j_3+j_4+2j_5+2j_6)}{2x_0(\Gg_1j_2+\Gg_2j_5)}
=-\frac{t_0(4\Gg_1+\Gg_2)}{x_0\Gg_1(\Gg_1+\Gg_2)}
\eeq{Stagg_6}
and this corresponds to the effective conductivity coefficient in the time direction $-\Gb_*$.

To derive the effective conductivity coefficient in the space direction, $\Ga_*$, we consider the current distribution illustrated here in \fig{8}, where, as before, we assign the convention
that currents flowing in the direction of positive time have positive sign, and currents flowing in the direction of negative time have negative sign. The currents
are still labelled $j_1$, $j_2$, $j_3$, $j_4$, $j_5$, and $j_6$ on the lines in \fig{8}.
Periodicity ensures the current flow has this structure. By conservation of current at the interfaces between the phases, we require that
\beq j_1+j_2=j_5+j_6, \quad j_3=-j_6, \quad j_4=-j_5. \eeq{Stagg_7} 
We let $V_i$, $i=1,2,\ldots, 12$ denote the potentials in each of the regions $1,2,\ldots, 12$ and we let $V_i'$, $i=1,3,5,6,\ldots,12$ denote the potentials in each of the regions
$1,3,5,6,\ldots,12$. Once again, for simplicty, we set $V_1'=0$. Then, from the rule that the potential jump across a characteristic line is $\Gg_i$ times the current flowing through it, we have
\beq\begin{aligned}
	&V_2 = \Gg_2j_5, \quad V_3=V_5=\Gg_2(j_5-j_6), \quad V_3'=V_5'=0,\quad V_4=-\Gg_1j_1+\Gg_2(j_5-j_6)=-\Gg_1j_2, \\&
	V_6 = \Gg_1j_1+\Gg_2(j_5-j_6),\quad V_6'=\Gg_1j_2,\quad V_7=\Gg_1(j_1+j_3)+\Gg_2(j_5-j_6),\quad V_7'=\Gg_1(j_2+j_4), \\& 
	V_8 = \Gg_1(j_1+j_3)+\Gg_2(j_5-2j_6), \quad V_8'=\Gg_1(j_2+j_4)-\Gg_2j_5, \quad V_9=\Gg_1(j_1+j_3)+2\Gg_2(j_5-j_6),\\& V_{10}=\Gg_1j_3+\Gg_2(j_5-j_6),\quad V_{10}'=\Gg_1j_4,
	\quad V_{11}=\Gg_1(2j_1+j_3)+\Gg_2(j_5-j_6),\\&V_{11}'=\Gg_1(2j_2+j_4),\quad 
	V_{12}=\Gg_1(j_1+j_3)+\Gg_2(j_5-j_6),\quad V_{12}'=\Gg_1(j_2+j_4).
\end{aligned}
\eeq{Stagg_8}
So in addition to \eq{Stagg_7} the currents must satisfy the constraint
\beq \Gg_1(j_1-j_2)=\Gg_2(j_5-j_6). \eeq{Stagg_9}
Therefore, we can choose $j_1$, and $j_2$ independently, and in terms of these
\beqa  j_4=-j_5& = &-\frac{1}{2}(j_1+j_2)+\frac{\Gg_1}{2\Gg_2}(j_2-j_1), \quad \nonum
j_3=-j_6& = &-\frac{1}{2}(j_1+j_2)-\frac{\Gg_1}{2\Gg_2}(j_2-j_1).
\eeqa{Stagg_10}
As $V_5'=V_1'$ there is no average electric field in the vertical direction. Similarly, there is no current field in the vertical direction. In the horizontal direction,
by looking at the current flux out of the unit cell, one sees the average current to the right is $(j_2+j_6-j_1-j_5)/t_0$. The average electric field pointing to the right
is $(V_{11}'-V_{11})/x_0$. The ratio of the average current field divided by the average electric field is then
\beq \frac{(j_2+j_6-j_1-j_5)/t_0}{(V_{11}'-V_{11})/x_0}=
\frac{x_0(\Gg_1+\Gg_2)}{t_0\Gg_1(3\Gg_2+\Gg_1)},
\eeq{Stagg_11}
and this corresponds to the effective conductivity coefficient in the space direction, $\Ga_*$. 

Therefore, the "effective conductivity tensor" for the staggered dynamics array is given by
\beq \BGs_*\!=\!\bpm \Ga_*\!\! & 0 \\ 0 & \!\!-\Gb_* \epm\!
=\! \bpm \frac{x_0(\Gg_1+\Gg_2)}{t_0\Gg_1(\Gg_1+3\Gg_2)}\!\! & 0\\ 0 &\!\! -\frac{t_0(4\Gg_1+\Gg_2)}{x_0\Gg_1(\Gg_1+\Gg_2)} \epm\!
=\!\bpm \frac{c_1(c_1+3c_2)(\Gg_1+\Gg_2)}{\Gg_1(c_1+c_2)(3\Gg_2+\Gg_1)}\!\!\!\!& 0 \\ 0 &\!\!\!\! -\frac{(c_1+c_2)(4\Gg_1+\Gg_2)}{\Gg_1c_1(c_1+3c_2)(\Gg_1+\Gg_2)}\epm.
\eeq{Stagg_12}
Then, the associated "effective speed of propagation" of the waves is 
\beq c_*=\sqrt{\Ga_*/\Gb_*}=\frac{c_1(c_1+3c_2)(\Gg_1+\Gg_2)}{(c_1+c_2)\sqrt{(3\Gg_2+\Gg_1)(4\Gg_1+\Gg_2)}}.
\eeq{Stagg_13}
Once again, this does not approach the wave velocity $c_1$, even when the parameters of phase $2$ approach those of phase $1$. If we had chosen 
our units of space and time so that $x_0=1$ and $t_0=1$ then the corresponding dimensionless speed would be
\beq \frac{c_*t_0}{x_0}=\frac{\Gg_1+\Gg_2}{\sqrt{(3\Gg_2+\Gg_1)(4\Gg_1+\Gg_2)}}.
\eeq{Stagg_14}

Again, we emphasize that we have put quotes around "effective conductivity tensor" and "effective speed"  because at this stage it is unclear what is their physical significance. As remarked in the
main text, if the conductivity tensor field
$\BGs(\Bx)$ had a very tiny imaginary part (as relevant to composites of hyperbolic materials, when $\BGs(\Bx)$ is replaced by the dielectric tensor field $\BGve(\Bx)$, and time is replaced
by a spatial variable) then indeed $\BGs_*$ would be the appropriate "speed" giving the effective "characteristic lines" of propagation. In the absence of such an imaginary part one needs
to derive the appropriate homogenized equations. This homogenized equation is unlikely to be simply $\Div\BGs_*\Grad \underline{V}=0$ where $\underline{V}(\Bx)$ is some coarse grained potential,
as it is quite evident from the branching nature of field patterns that there should be some term giving dispersion in the effective equations.

\begin{figure}[!ht]
	\centering
	\includegraphics[width=0.7\textwidth]{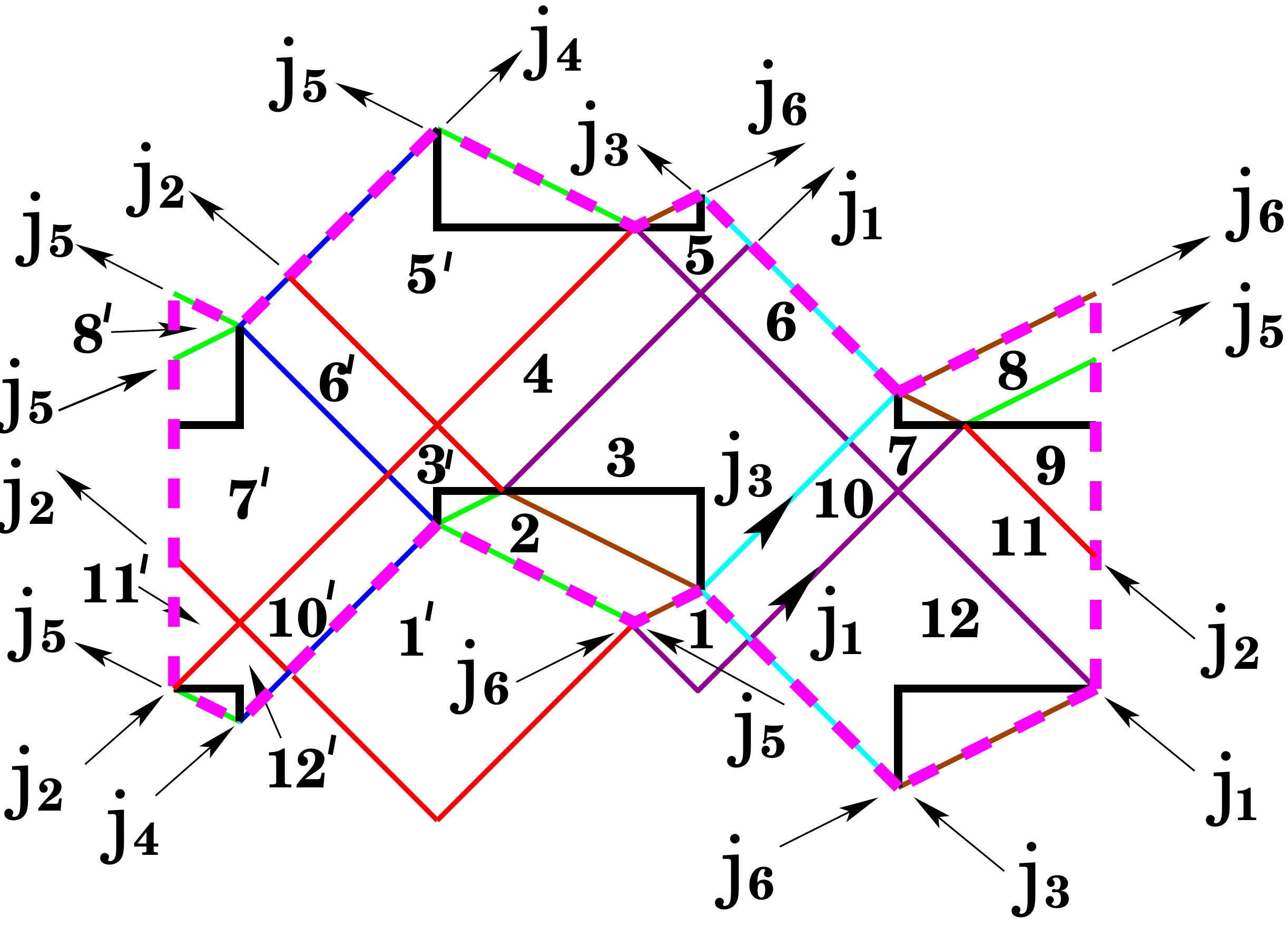}
	\caption{The current flows in the unit cell for the staggered array dynamics with the average electric and current fields 
		directed along the time axis.}
	\labfig{7}
\end{figure}

\begin{figure}[!ht]
	\centering
	\includegraphics[width=0.7\textwidth]{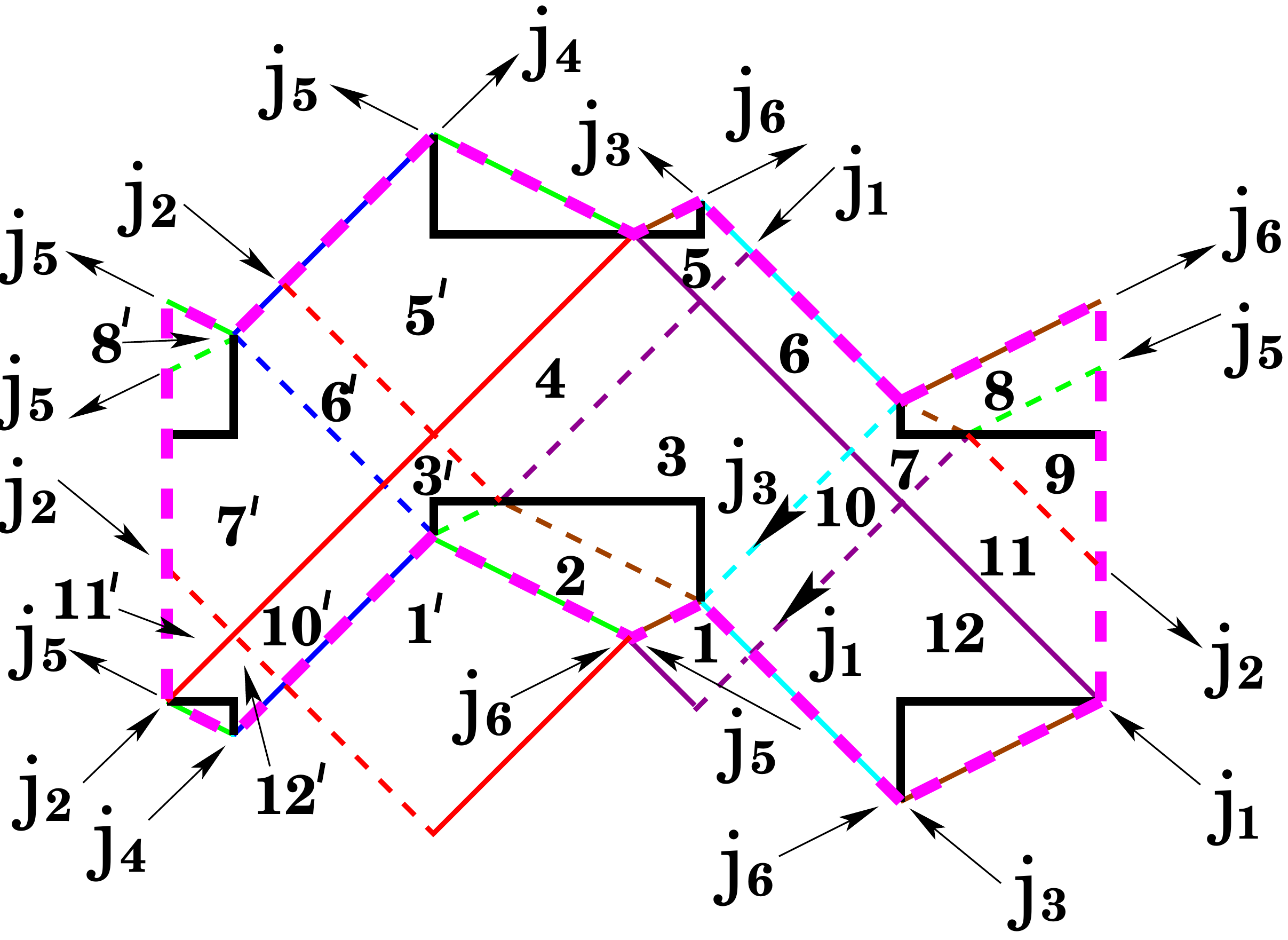}
	\caption{The current flows in the unit cell for the staggered array dynamics with the average electric and current fields 
		directed along the space axis. Here the dashed lines denote currents in the negative time direction, i.e. a current $j_1$ flowing in the negative time direction
		is equivalent to a current $-j_1$ flowing in the positive time direction.}
	\labfig{8}
\end{figure}

We tested the theoretical results obtained here by using the same procedure adopted for the aligned dynamics array in Section 5 of the main text. In particular, by choosing as test points the points of intersection between the characteristics and the horizontal line $t=\tau$, as in \fig{green}, we obtain 8 test points. As before, we calculate, as detailed below, the Green function that allows one to determine the transfer matrix that gives the distribution of currents at a certain time $t=\tau+n\,t_0$ with $n$ fixed, given the currents at time $t=\tau+(n-1)\,t_0$.

\section{Appendix C: Green's function for the space-time staggered inclusion microstructure}
\setcounter{equation}{0}

By using the notation introduced in Section 6 of the main text, the components of the Green function related to the staggered geometry are given step by step by injecting current at the following points: 
\beq\nonumber
\begin{aligned}
	&\text{Point 1 of cell 0}: j(1,2,0)=1\quad\Rightarrow\quad G(4,1,0)=1\\&
	\text{Point 2 of cell 0}: j(2,2,0)=1\quad\Rightarrow\quad G(7,2,-1)=1; \, G(1,2,0)=\frac{\Gg_1-\Gg_2}{\Gg_1+\Gg_2};\,G(3,2,0)=-\frac{\Gg_1-\Gg_2}{\Gg_1+\Gg_2}\\&
	\text{Point 3 of cell 0}: j(3,2,0)=1\quad\Rightarrow\quad\left\{ \begin{aligned}
		&G(8,3,-1)=-\frac{\Gg_1-\Gg_2}{\Gg_1+\Gg_2};
		\,G(1,3,0)=-\frac{(\Gg_1-\Gg_2)^2}{(\Gg_1+\Gg_2)^2};\\&\,G(2,3,0)=\frac{\Gg_1-\Gg_2}{\Gg_1+\Gg_2};\,G(3,3,0)=\frac{(\Gg_1-\Gg_2)^2}{(\Gg_1+\Gg_2)^2};\,G(6,3,0)=1\end{aligned}\right.\\&
	\text{Point 4 of cell 0}: j(4,2,0)=1\quad\Rightarrow\quad\left\{ \begin{aligned}
		&G(8,4,-1)=-\frac{\Gg_1-\Gg_2}{\Gg_1+\Gg_2};
		\,G(1,4,0)=-\frac{(\Gg_1-\Gg_2)^2}{(\Gg_1+\Gg_2)^2};\\&\,G(2,4,0)=\frac{\Gg_1-\Gg_2}{\Gg_1+\Gg_2};
		\,G(3,4,0)=\frac{(\Gg_1-\Gg_2)^2}{(\Gg_1+\Gg_2)^2};\\&\,G(7,4,0)=-\frac{\Gg_1-\Gg_2}{\Gg_1+\Gg_2};
		\,\,G(8,4,0)=\frac{\Gg_1-\Gg_2}{\Gg_1+\Gg_2};\,G(1,4,1)=1 \end{aligned}\right.\\&
	\text{Point 5 of cell 0}: j(5,2,0)=1\quad\Rightarrow\quad \left\{\begin{aligned} &G(8,5,-1)=1; \, G(1,5,0)=\frac{\Gg_1-\Gg_2}{\Gg_1+\Gg_2};\,G(3,5,0)=-\frac{\Gg_1-\Gg_2}{\Gg_1+\Gg_2};\\&\,G(6,5,0)=\frac{\Gg_1-\Gg_2}{\Gg_1+\Gg_2};\,G(7,5,0)=\frac{(\Gg_1-\Gg_2)^2}{(\Gg_1+\Gg_2)^2};\\&\,G(8,5,0)=-\frac{(\Gg_1-\Gg_2)^2}{(\Gg_1+\Gg_2)^2};\,G(1,5,1)=-\frac{\Gg_1-\Gg_2}{\Gg_1+\Gg_2}\end{aligned}\right.\\&
	\text{Point 6 of cell 0}: j(6,2,0)=1\quad\Rightarrow\quad G(7,6,0)=-\frac{\Gg_1-\Gg_2}{\Gg_1+\Gg_2}; \, G(8,6,0)=\frac{\Gg_1-\Gg_2}{\Gg_1+\Gg_2};\,G(3,6,1)=1\\&
	\text{ Point 7 of cell 0}: j(7,2,0)=1\quad\Rightarrow\quad \left\{\begin{aligned} &G(2,7,0)=1; \, G(6,7,0)=\frac{\Gg_1-\Gg_2}{\Gg_1+\Gg_2};\,G(7,7,0)=\frac{(\Gg_1-\Gg_2)^2}{(\Gg_1+\Gg_2)^2};\\&\,G(8,7,0)=-\frac{(\Gg_1-\Gg_2)^2}{(\Gg_1+\Gg_2)^2};\,G(1,7,1)=-\frac{\Gg_1-\Gg_2}{\Gg_1+\Gg_2}\end{aligned}\right.\\&
	\text{Point 8 of cell 0}: j(8,2,0)=1\quad\Rightarrow\quad G(5,8,0)=1\\
\end{aligned}
\eeq{Green}

Obviously, all the other components are equal to zero.

The eigenvalues of the transfer matrix, defined as $T_{(k,m),(k,'m')}=G_{k,k'}(m-m')$, like in Section 6 of the main text, are plotted in \fig{eigen_stag} in the case $\Gg_1=1$ and $\Gg_2=3$. Here we consider 10 cells, for a total of 80 points (8 points for each unit cell) so that the number of eigenvalues is equal to 80 (we write the transfer matrix as a $80\times 80$ matrix that, applied to a vector with 80 components describing the current distribution at the 80 points at a certain time, provides the vector with 80 components representing the currents at the same 80 points after a cycle of time). 

\begin{figure}[!ht]
	\centering
	\includegraphics[width=0.7\textwidth]{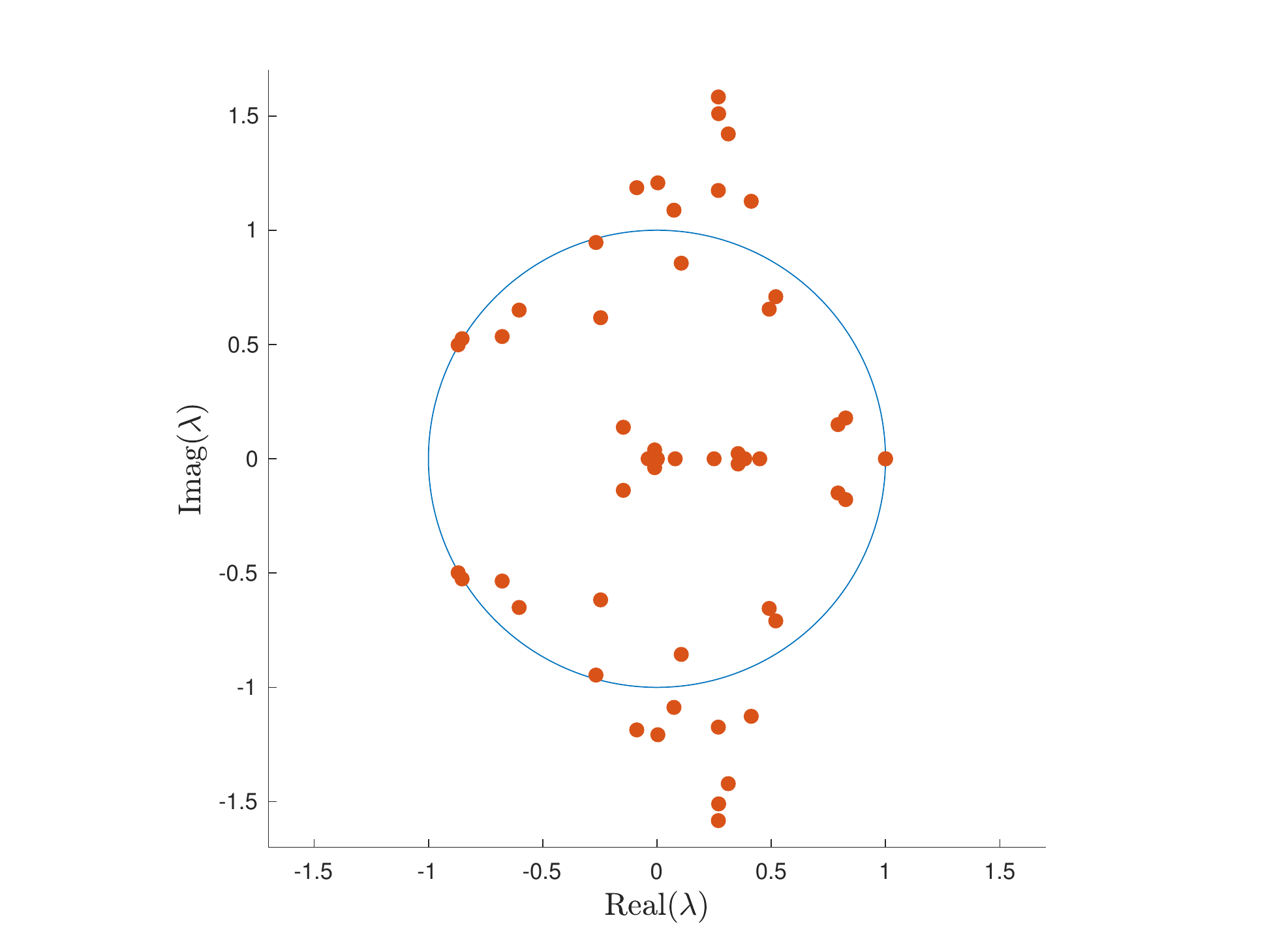}
	\caption{Eigenvalues of the transfer matrix in the case $\Gg_1=1$ and $\Gg_2=3$.}
	\labfig{eigen_stag}
\end{figure}

\end{document}